\documentclass[aps,prb,twocolumn,superscriptaddress,showpacs,floatfix]{revtex4}
\usepackage[american]{babel}
\usepackage{graphicx}
\usepackage{amsmath}
\usepackage{amssymb}
\usepackage{hyperref}
\newcommand{\rem}[1]{}

\begin{document}

\title{Soliton dynamics in a solid lubricant during sliding friction}

\author{Anna Vigentini}
\affiliation{Dipartimento di Fisica, Universit\`a degli Studi di Milano,
Via Celoria 16, 20133 Milano, Italy
}

\author{Barbara Van Hattem}
\affiliation{Dipartimento di Fisica, Universit\`a degli Studi di Milano,
Via Celoria 16, 20133 Milano, Italy
}

\author{Elena Diato}
\affiliation{Dipartimento di Fisica, Universit\`a degli Studi di Milano,
Via Celoria 16, 20133 Milano, Italy
}

\author{Paolo Ponzellini}
\affiliation{Dipartimento di Fisica, Universit\`a degli Studi di Milano,
Via Celoria 16, 20133 Milano, Italy
}

\author{Tommaso Meledina}
\affiliation{Dipartimento di Fisica, Universit\`a degli Studi di Milano,
Via Celoria 16, 20133 Milano, Italy
}

\author{Andrea Vanossi}
\affiliation{CNR-IOM Democritos National Simulation Center, Via Bonomea
  265, 34136 Trieste, Italy}
\affiliation{International School for Advanced Studies (SISSA), Via Bonomea
  265, 34136 Trieste, Italy}

\author{Giuseppe Santoro}
\affiliation{International School for Advanced Studies (SISSA), Via Bonomea
  265, 34136 Trieste, Italy}
\affiliation{CNR-IOM Democritos National Simulation Center, Via Bonomea
  265, 34136 Trieste, Italy}
\affiliation{International Center for Theoretical Physics (ICTP), Strada
  Costiera 11, 34151 Trieste, Italy}

\author{Erio Tosatti}
\affiliation{International School for Advanced Studies (SISSA), Via Bonomea
  265, 34136 Trieste, Italy}
\affiliation{CNR-IOM Democritos National Simulation Center, Via Bonomea
  265, 34136 Trieste, Italy}
\affiliation{International Center for Theoretical Physics (ICTP), Strada
  Costiera 11, 34151 Trieste, Italy}

\author{Nicola Manini}
\affiliation{Dipartimento di Fisica, Universit\`a degli Studi di Milano,
Via Celoria 16, 20133 Milano, Italy
}
\affiliation{CNR-IOM Democritos National Simulation Center, Via Bonomea
  265, 34136 Trieste, Italy}
\affiliation{International School for Advanced Studies (SISSA), Via Bonomea
  265, 34136 Trieste, Italy}

\date{\today}

\begin{abstract}
Recent highly idealized model studies of lubricated nanofriction for two
crystalline sliding surfaces with an interposed thin solid crystalline
lubricant layer showed that the overall relative velocity of the lubricant
$v_{\rm lub} / v_{\rm slider}$ depends only on the ratio of the lattice
spacings, and retains a strictly constant value even when system parameters
are varied within a wide range.
This peculiar ``quantized'' dynamical locking was understood as due to the
sliding-induced motion of misfit dislocations, or soliton structures.
So far, the practical relevance of this concept to realistic sliding three
dimensional crystals has not been demonstrated.
In this work, by means of classical molecular dynamics simulations and
theoretical considerations, we realize a realistic three-dimensional
crystal-lubricant-crystal geometry.
Results show that the flux of lubricant particles associated with the
advancing soliton lines gives rise here too to a quantized velocity
ratio.
Moreover, depending on the interface lattice spacing mismatch, both forward
and backward quantized motion of the lubricant is predicted.
The persistence under realistic conditions of the dynamically pinned state
and quantized sliding is further investigated by varying sliding speed,
temperature, load, and lubricant film thickness.
The possibilities of experimental observation of quantized sliding are also
discussed.
\end{abstract}

\maketitle

\section{Introduction}

The problem of boundary lubricated friction of two perfect sliding crystal
surfaces is fascinating both from the fundamental point of view and for
applications in the wider context of nanofriction.\cite{VanossiRMP13}
Intriguing and unexpected behavior of the relative lubricant velocity
have recently been reported in numerical simulations, depending on the
``degree'' of geometrical incommensurability defining the moving interface.
The main nontrivial feature is the asymmetry in the sliding velocity of the
intermediate lubricant sheet relative to the two
substrates.\cite{Vanossi06, Santoro06, Cesaratto07, Vanossi07Hyst,
  Manini07extended, Vanossi08TribInt, Manini07PRE, Vanossi07PRL,
  Manini08Erice, Castelli09, Castelli08Lyon}.
Moreover, and even more strikingly, the lubricant mean velocity takes a
constant, ``quantized'', value uniquely determined by the
incommensurability ratios of the three spatial periodicities involved --
the two sliders and the interposed solid lubricant -- and is insensitive to
other physical parameters of the model.
The sliding steady state versus overall sliding velocity, as well as other
parameters, is characterized by perfectly flat plateaus in the ratio of the
time-averaged lubricant center of mass (c.m.) velocity to the externally
imposed relative speed $v_{\rm ext}$ of the two sliders.
This amounts to a kind of ``dynamical incompressibility'' or dynamic
pinning, namely, identically null velocity response to perturbations or
fluctuations trying to deflect the relative lubricant velocity away from
its quantized value.
The occurrence of this surprising regime of motion was ascribed to the
intrinsic topological nature of this locked dynamics.
This phenomenon, investigated in detail in rather idealized one-dimensional
(1D) geometries \cite{Vanossi06, Santoro06, Cesaratto07, Vanossi07Hyst,
  Manini07extended, Vanossi08TribInt, Manini07PRE, Vanossi07PRL,
  Manini08Erice}, was explained by the grip exerted by one slider onto the
topological solitons (called kinks or antikinks in one-dimension) that the
embedded solid lubricant lattice forms with the other slider.
The pinning of these solitons by the first slider causes their rigid
dragging at the full sliding speed $v_{\rm ext}$.
As a result the overall mean lubricant speed is a fixed ratio $w$ of the
slider's speed, strictly determined by the soliton spatial density, a
purely geometrical factor $|w|<1$.
Simulation evidence of this particular sliding regime was also confirmed
for a less idealized 1+1-dimensional (1+1D) model of boundary lubrication
\cite{Castelli09,Castelli08Lyon}, where Lennard-Jones (LJ) interacting
atoms were allowed to move freely, parallel and perpendicularly to the
sliding direction.
Solitons formed in this case too, and their influence transmitted from one
slider to the other across the lubricant film even when the thickness is as
large as six atomic layers.

In this work, we simulate lubricated sliding in a fully 3D prototypical
model.
We again find that, under fairly general conditions, the lubricant slides
relative to a fixed surface with a mean relative lubricant velocity
component in the driving direction $w=v_{{\rm c.m.}\,x}/v_{\rm ext}$, which
is ``quantized'' to a basically parameter-independent value $w= w_{\rm
  quant}$, much as was observed for the essentially 1D models.
Confirming its soliton nature here too, we characterize the properties and
limitations of the quantized-velocity dynamics in the 3D model, showing
that the quantized sliding is robust against wide-range variations of
different model parameters.

\begin{figure}
\centering
\includegraphics[width=85mm,angle=0,clip=]{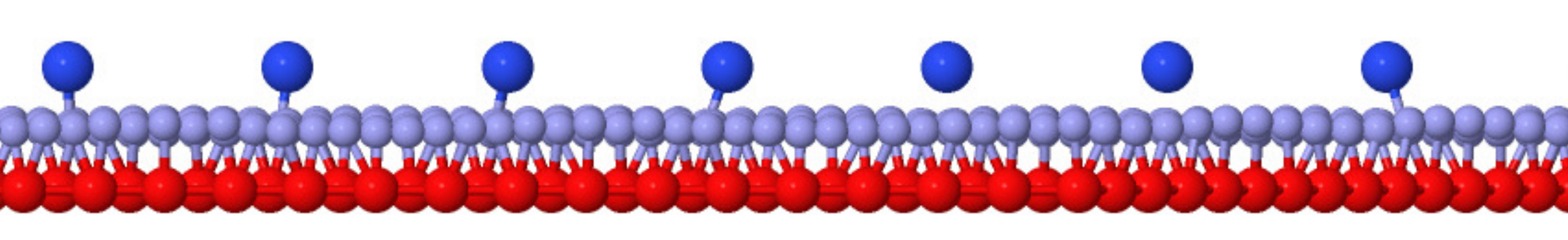}
\includegraphics[width=85mm,angle=0,clip=]{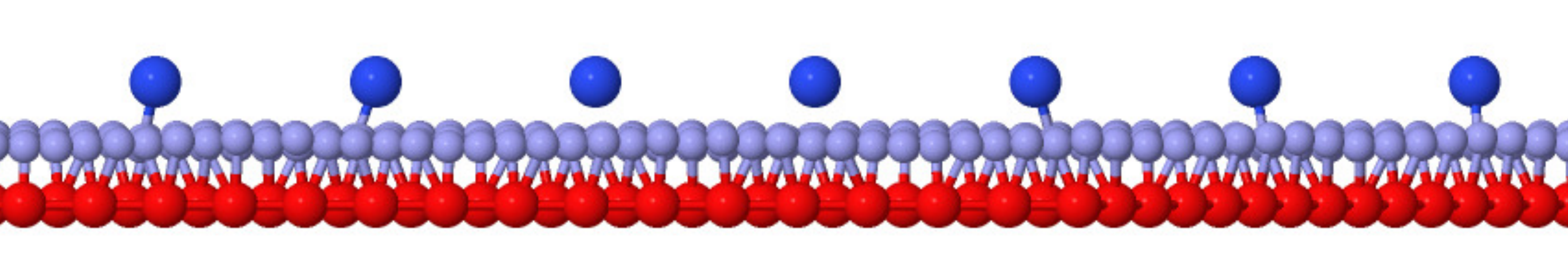}
\includegraphics[width=85mm,angle=0,clip=]{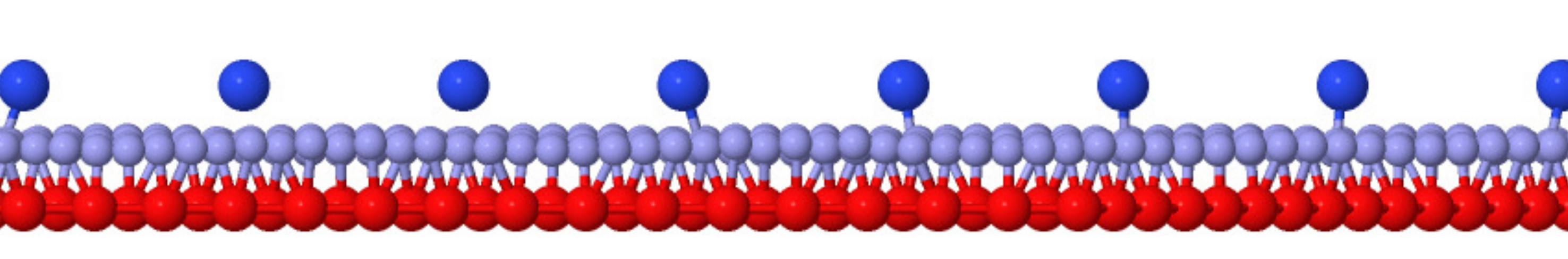}
\caption{\label{snap:fig} (Color online)
Side view of the substrate-lubricant-substrate sandwich, with the static
bottom substrate (red), the mobile lubricant atoms (light blue, smaller)
and the top rigid slider, with much larger spacing (dark blue).
Three successive time frames illustrate the ``caterpillar'' soliton motion
driven by the rightward advancing top layer.
Note the small vertical corrugations of the lubricant layer.
}
\end{figure}

An intuitive and suggestive picture of the advancing solitons in the
quantized state can be appreciated by the side view of the 3D geometry of
Fig.~\ref{snap:fig}.
One can note the characteristic ``caterpillar'' motion executed by the
lubricant particles in contact with the closest-matched crystal surface.
In the 3D geometry actually the soliton (Moir\'e) pattern is a 2D feature,
which in general implies additional characteristics, such as mismatches
induced by relative lattice rotation \cite{Novaco77}.
In this 3D study we will however restrict our investigation to mutually
aligned incommensurate geometries, deferring the rotated cases to future
work.

\section{The model}\label{model:sec}

We represent the two confining 3D crystal surfaces by perfectly periodic 2D
($xy$) monolayers, rigidly arranged in close-packed triangular lattices
representing, e.g. the (111) face of a cubic crystal. Between these two
rigid planar sliders we insert $N_{\rm layer}$ layers of generally
crystallized but mobile lubricant atoms, see Fig.~\ref{snap:fig}.
Each layer is composed of point-like classical particles of unit mass ($m =
1$).
While the reciprocal intra-layer positions of top and bottom slider atoms
are of course fixed, the atoms composing the lubricant film
move freely under the action of pairwise (6,12) LJ
interactions among one another and with the rigid atoms forming the top and
the bottom surfaces.
The standard LJ interaction
\begin{equation}\label{eq_LJ}
\mathcal \phi _{\rm LJ} (r) =\varepsilon \left[\left(\frac{\sigma}{r}\right)^{12}-2\left(\frac{\sigma}{r}\right)^6 \right]
\,,
\end{equation}  
is characterized by depth $\varepsilon $ and radius $r= \sigma$.
We truncate the interaction at a cutoff radius $R_{\rm C} =2.5 \sigma$ and
shift the 2-body potential energy to eliminate the energy discontinuity as
follows:
\begin{equation}\label{eq_Cutoff}
\mathcal \phi (r)=
\begin{cases}
\phi_{\rm LJ}(r)-\phi_{\rm LJ}(R_{\rm C}) & \text{ $r \leq $$R_{\rm C}$} \\
0 & \text{ $r > $$R_{\rm C}$}
\end{cases}
\,.
\end{equation}

The motion of the $j$-th lubricant particle is ruled by the equation of
motion
\begin{equation}\label{eq_motion1}
\begin{split}
m\ddot{\vec{r}}_{j} = &- \sum_{i_{\rm t} = 1}^{N_{\rm t}}
\frac{\partial}{\partial{\vec{r}_{j}}}\phi^{\rm t,\rm p}(|
\vec{r}_{j} - \vec{r}_{i_{\rm t}}|) +
\\
& - \sum_{{j'=1} \atop {j' \neq j}}^{N_{\rm p}}
\frac{\partial}{\partial{\vec{r}_{j}}}
\phi^{\rm p,\rm p}(| \vec{r}_{j} - \vec{r}_{j'}|)+\\
    &- \sum_{i_{\rm b} = 1}^{N_{\rm b}}
\frac{\partial}{\partial{\vec{r}_{j}}}
\phi^{\rm b,\rm p}(| \vec{r}_{j} - \vec{r}_{i_{\rm b}}|)
+\vec{f}_{{\rm damp}\, j} + \vec{f}_{j}(t)
\,,
\end{split}
\end{equation}
where $\vec r_j$ is the position of the $j$-th lubricant particle;
$\vec{r}_{i_{\rm t}} $ and $\vec{r}_{i_{\rm b}} $ are the positions of the
top and bottoms slider atoms, $N_{\rm b} $, $N_{\rm p} $ and $N_{\rm t}$
are the numbers of the bottom, lubricant and top particles, and $\phi^{\rm
  b,\rm p} $, $\phi^{\rm p,\rm p} $ and $\phi^{\rm t,\rm p} $ are the
truncated 2-body potential energies for the interactions between
bottom-lubricant, lubricant-lubricant and top-lubricant particles,
respectively, characterized by generally different $\sigma $ and
$\varepsilon$ parameters, as specified below.
$\vec{f}_{{\rm damp}\, j}$ and $\vec{f}_{j}(t)$ are a damping force and a
random force respectively, used to implement a Langevin dynamics, as
detailed below.

By convention, we select the bottom slider as our reference frame.
The top slider is forced to move rigidly along $\hat x$ at a fixed
horizontal velocity $\dot{r}_{x_{i_{\rm t}}}^{\rm top}(t) \equiv v_{\rm
  ext}$, under an external downward force $-F_{\rm load}\,\hat{z}$ applied
to each particle in the slider.
It also generally moves along the $\hat{y} $ and $\hat{z} $ axes (its
inertia equals the total mass $N_{\rm t}$ of its atoms) under the
interaction between its atoms and those of the lubricant film.
For these $\hat y$ and $\hat z$ components, the motion of the top slider is
described by
\begin{eqnarray}\label{eq_topdynamics1}
N_{\rm t}m\ddot{r}_{y_{i_{\rm t}}}^{\rm top}(t)
&=& -  \sum_{i'_{\rm t}=1}^{N_{\rm t}}\sum_{j=1}^{N_{\rm p}}
\frac{\partial}{\partial{r_y}}
\phi^{\rm t,\rm p}(|\vec r_{i'_{\rm t}}-\vec r_{j}|) 
+F_{{\rm th}\, y}
\,,
\\\label{eq_topdynamics2}
N_{\rm t}m\ddot{r}_{z_{i_{\rm t}}}^{\rm top}(t)
&=& - \sum_{i'_{\rm t}=1}^{N_{\rm
    t}}\sum_{j=1}^{N_{\rm p}}\frac{\partial}{\partial{r_z}}
\phi^{\rm t,\rm p}(|\vec{r}_{i'_{\rm t}}-\vec{r}_{j}|)
\\\nonumber
&& +F_{{\rm th}\, z}-N_{\rm t}F_{\rm load}
\,,
\end{eqnarray}
where the components of the thermostat force $\vec F_{\rm th}$ are
discussed below.
As all equations for $r_{y/z_{i_{\rm t}}}^{\rm top}$ are the same,
irrespective of $i_{\rm t} $, in practice their solution only differs by a
translation ${\vec r}_{_{i_{\rm t}}}^{\rm top} \equiv {\vec r}^{\rm top} +
{\vec r}_{i_{\rm t}}^{\rm init}$ (where ${\vec r}_{i_{\rm t}}^{\rm init}$
are the initial positions of the rigid top 2D lattice), so that equations
for $r_y^{\rm top}$ and $r_z^{\rm top}$ only are integrated.

\subsection{Frictional work and thermostat}

The total force needed to maintain the top slider at the fixed velocity
$v_{\rm ext}$ compensates exactly the total force which the lubricant
exerts on the top slider itself:
\begin{equation}\label{Eq_Ffric}
F_{\rm frict}=
 \sum_{i'_{\rm t}=1}^{N_{\rm t}}\sum_{j=1}^{N_{\rm p}}
\frac{\partial}{\partial{r_x}}
\phi^{\rm t,\rm p}(|\vec r_{i'_{\rm t}}-\vec r_{j}|) 
-F_{{\rm th}\, x}
\,.
\end{equation}
The work of this frictional force
\begin{equation}\label{eq_Ediss}
W_{\rm frict}=
\int_0^\tau \!F_{\rm frict} v_{\rm ext}\,dt =
v_{\rm ext}\int_0^\tau \! F_{\rm frict}\,dt =
\tau v_{\rm ext}\bar F_{\rm frict}
\end{equation}
represents the total Joule heat that the advancing top layer pumps into the
mechanical system over a time interval $\tau$.

To remove this Joule heat, to reach a steady state, and to control the rise
of lubricant temperature in this driven system, we use a standard
implementation of the Langevin dynamics, Eq.~\eqref{eq_motion1}, including
a phenomenological viscous damping term, plus a Gaussian random force
$\vec{f_{j}}(t)$.
To avoid biasing the lubricant motion by privileging either the bottom or
the top reference frame, the damping force includes two contributions
representing the energy dissipation into both sliders
\begin{equation}\label{eq_Langevin}
\vec{f}_{{\rm damp}\,j} =
- \eta \dot{\vec{r}}_{j}- \eta(\dot{\vec{r}}_{j}-\dot{\vec{r}}_{\rm t})
\,.
\end{equation}
Taking into account this twofold contribution to dissipation, the
zero-average Gaussian random forces satisfy
\begin{equation}\label{eq_Gaussian}
\langle f_{j\beta}(t)f_{j'\beta'}(t')\rangle =
 4\eta k_{\rm B}T\delta_{jj'}\delta_{\beta\beta'}\delta(t-t')
\,,
\end{equation}
(with $\beta,\beta'=x,y,z$ components), so that in a non-sliding regime
($v_{\rm ext}=0$) the Langevin thermostat leads to a stationary state
characterized by standard Boltzmann equilibrium average kinetic energy of
the lubricant:
\begin{equation}\label{eq_Ekin}
\langle E_{\rm k} \rangle = 3N_{\rm p}\frac{1}{2}k_{\rm B}T
\,.
\end{equation}

The damping force contribution representing the energy dissipation into the
top slider requires a force balance (Newton's third law) term in
Eqs.~\eqref{eq_topdynamics1} and \eqref{eq_topdynamics2} for the top layer:
\begin{equation}\label{eq_addforce}
\vec F_{\rm th}
=
\eta \sum_{i}^{N_{\rm p}}(\dot{\vec{r}}_{i}-\dot{\vec{r}}_{\rm t}) =
\eta N_{\rm p}(\vec{v}_{\rm c.m.}-\dot{\vec{r}}_{\rm t})
\,.
\end{equation}
While the $\hat y$ and $\hat z$ components of this additional term have a
real influence on the top motion through Eqs.~\eqref{eq_topdynamics1} and
\eqref{eq_topdynamics2}, of course its $\hat x$ component does not.
It only contributes to the external force $F_{\rm frict}$ required to
maintain the top velocity $\hat x$ component constant and equal to $v_{\rm
  ext}$, with the last term in Eq.~\eqref{Eq_Ffric}.

As long as the value of $\eta$ is so small \cite{smalleta:comment} that it
produces an underdamped dynamics, the thermostat perturbs the atomistic
dynamics only marginally.
Under this condition, the Langevin method represents a simple but
numerically stable and effective phenomenological approach to describe energy
dissipation into the substrates occurring e.g., through the excitation of
phonons and (in the case of metals) of electron-hole pairs, etc.
We verified that all qualitative results are insensitive to the value of
$\eta$ (as long as it is small enough), although quantitative issues such
as the precise boundary of the quantized sliding regime do depend on
$\eta$.
More refined methods were proposed and adopted in similar simulations
\cite{Braun01a,Braun06, Kantorovich08a, Kantorovich08b, molshapeLin,
  Benassi10, BraunManini11, Benassi12} but to investigate the occurrence
and main properties of the quantized sliding phenomenon, a simple Langevin
approach to power dissipation is sufficient and appropriate.

\subsection{Length scales and units}

The sliding system involves three generally different solids, two sliders
and a lubricant, which in their crystalline state are characterized by
generally different lattice spacings: $a_{\rm p}$, $a_{\rm t}$, and $a_{\rm
  b}$.
For the particle-particle interaction inside the lubricant we take the LJ
radius $\sigma_{\rm pp} = 1.01 a_{\rm p}$ so as to compensate approximately
first-neighbor repulsion with second- and third-neighbor attraction.
%
Interactions within each of the rigid sliders are of course not needed.
However, one could still introduce them for convenience with radii
$\sigma_{\rm tt} = a_{\rm t}$ and $\sigma_{\rm bb}$ = $a_{\rm b}$, and fix
slider-lubricant interaction radii $\sigma_{\rm tp}$ and $\sigma_{\rm bp}$
e.g.\ by means of the Lorentz-Berthelot mixing rules \cite{Allen91}:
\begin{equation}\label{eq_sigmaLJ}
\sigma_{\rm tp} = \frac{1}{2} (\sigma_{\rm tt}+\sigma_{\rm pp})
\,,\quad
\sigma_{\rm bp} = \frac{1}{2} (\sigma_{\rm bb}+\sigma_{\rm pp})
\,.
\end{equation}
In practice however we fix the radii according to $\sigma_{\rm
  tp}=\sigma_{\rm bp}= 1.02 a_{\rm b}$, and for simplicity, we fix the same
interaction energy $\varepsilon_{\rm tp} = \varepsilon_{\rm pp} =
\varepsilon_{\rm bp} = \varepsilon $ for all pairwise coupling terms,
unless otherwise noted.

\begin{table}
\begin{center}
\begin {tabular}{c|c|c}
\hline
\hline
Physical quantity & Natural units & Typical value\\
\hline
length & $a_{\rm b}$ 					& $0.2$ nm \\
mass & $m$ 						& $50$ a.m.u.$\simeq
							8.3\times 10^{-26}$ kg\\
energy & $\varepsilon_{\rm pp}$ 			& $1$ eV $\simeq
							1.6\times 10^{-19}$ J\\
time & $a_{\rm b}\,m^{1/2}\varepsilon_{\rm pp}^{-1/2}$	& $0.14$ ps \\
velocity $v$ & $ m^{-1/2} \varepsilon_{\rm pp}^{1/2}$	& $1400$ m/s \\
force & $a_{\rm b}^{-1} \, \varepsilon_{\rm pp}$	& $0.8$ nN \\
\hline
\hline
\end{tabular}
\end{center}
\caption{\label{units:tab}
Natural units for several mechanical quantities in a system where length,
mass and energy are measured in units of $a_{\rm b}$, $m$, $\varepsilon$.
Typical physical values are also indicated.
}
\end{table}

We consider a set of ``natural'' units in terms of $\varepsilon$ (energy),
$a_{\rm b}$ (length), and $m$ (mass).
All quantities are then expressed as dimensionless numbers.
To obtain a physical quantity in its explicit dimensional form, one should
multiply its simulated numerical value by the corresponding natural units
listed in Table~\ref{units:tab}.

The spacings $a_{\rm t}$, $a_{\rm p}$, and $a_{\rm b}$, and the angles of
relative rotation, define the initial conditions for the sliders and the
lubricant lattices.
Each atomic layer is initially a perfect 2D triangular lattice.
We stack complete layers, realizing an fcc crystalline film of lubricant as
it would be at low temperature.
The initial vertical separation between successive lubricant layers is of
the order of $\sqrt{2/3} a_{\rm p}$.

The three different spacings $a_{\rm t}$, $a_{\rm b}$ and $a_{\rm p}$ give
rise to two independent ratios affecting the 2D lattice mismatches:
\begin{equation}\label{eq_ratios}
r_{\rm t} = \frac{a_{\rm t}}{a_{\rm p}}
\,,\quad
r_{\rm b} = \frac{a_{\rm b}}{a_{\rm p}}
\,.
\end{equation}

We perform the numerical integration of Eqs.~\eqref{eq_motion1},
\eqref{eq_topdynamics1} and \eqref{eq_topdynamics2} by means of an adaptive
fourth-order Runge-Kutta-Fehlberg method, when $T = 0$, or, for finite $T$,
a 6-steps Runge-Kutta algorithm involving Langevin random forces,
Eq.~\eqref{eq_Gaussian}.

\subsection{Boundary conditions}\label{boundary:sec}

In order to explore with ease a large number of different configurations
and to follow their evolution long enough for the top and lubricant to
advance by several lattice spacings, our simulations involve a number of
lubricant atoms $N_{\rm p}\lesssim 10^3$, which is exceedingly small
compared to those involved in a realistic sliding interface (easily of the
order of $10^7$ in a $\mu$m$^2$).
To alleviate the effect of finite size and impose precise lattice-spacing
ratios, we use periodic boundary conditions (PBC) in the $xy$ plane: the
particles are enclosed in a supercell generated by two vectors
$\vec{a}_i^{\rm cell}$ of length $L$, replicated infinitely by means of
rigid translations.
Each particle $j$ in the box interacts not just with the other particles
$j'$ in the supercell, but also with their translated images in the nearest
neighboring cells by means of a standard minimum-image algorithm
\cite{Allen91}.
In the third ($\hat z$) direction, the lubricant is of course confined by
top and bottom sliders.
In the simple case in which the crystalline directions of the bottom,
lubricant, and top lattices are parallel, it is straightforward to
construct the appropriate supercell, whose side $L$ is an integer multiple
(e.g.\ the smallest multiple) of all three 2D lattice spacings, which have
therefore to be taken mutually commensurate.
For example, for $a_{\rm b}=1$, $a_{\rm p}=25/29$, $a_{\rm t}=25/4$, the
smallest supercell is obtained by taking $L=25$.

\subsection{The coverage ratio}

The quantized velocity state was interpreted in 1D as the dynamical pinning
of the periodic soliton pattern on the comparably long-wavelength
corrugation potential produced by the top substrate \cite{Vanossi06,
  Vanossi07PRL}.
Isomorphic to a static depinning transition (the role of particles now
taken by the moving kinks of the lubricant-substrate interface), although
different in nature, this pinning should be particularly robust for perfect
one-to-one commensurate matching of the inter-soliton spacing $a_{\rm sol}$
and the top-slider lattice spacing $a_{\rm t}$,\cite {Vanossi07PRL} a
condition where the soliton dragging should be especially effective in
producing the quantized state.

Whenever the top lattice and the soliton pattern are aligned along the same
crystalline directions, it makes sense to define a length ratio
\begin{equation}\label{eq_coverage}
\Theta = \sqrt{\frac{N_{\rm sol}}{N_{\rm t}}} =
\frac{a_{\rm t}}{a_{\rm sol}}
\,,
\end{equation}
defining a ``coverage'', and whose actual value depends on the spacing of
solitons $a_{\rm sol}$.
The latter in turn is tuned by the geometric mismatch condition between the
lubricant and bottom layers, as detailed in Sect.~\ref{qss:sec}.

For most of the simulations described in the following we have selected an
appropriate $r_{\rm b}$ to obtain $\Theta=1$.
However, as discussed later, we also investigated the quantized sliding for
the less specific geometrical configuration when $r_{\rm b}$ is such that
$\Theta$ deviates from unity.

\section{Results}\label{Results:sec}

\begin{figure}
\centerline{
\includegraphics[width=85mm,angle=0,clip=]{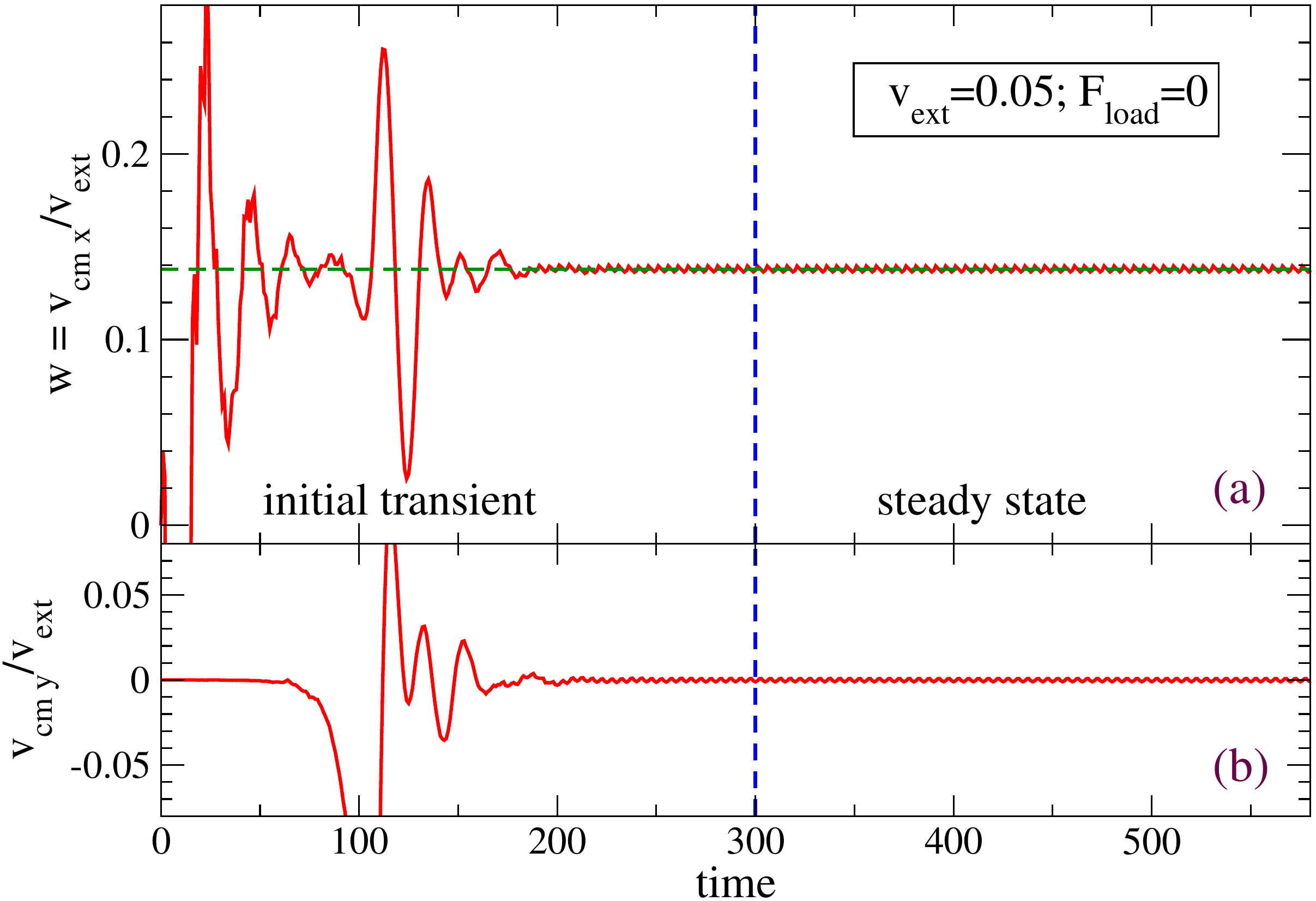}
}
\caption{\label{transient:fig} (Color online)
A typical approach to the steady state for the $N_{\rm layer}=1$ model
represented in Fig.~\ref{modelsketch}, with unrotated layers characterized
by $a_{\rm t} =25/4 =6.25$, $a_{\rm p} =25/29$, $a_{\rm b}=1$.
(a) Average lubricant velocity component in the driving direction,
$w=v_{{\rm c.m.}\,x}/v_{\rm ext}$ (normalized by the top externally-fixed
speed), as a function of time.
After an initial transient, $w$ starts to fluctuate around the value
predicted by Eq.~\eqref{w_quant}: $w_{\rm quant}= 4/29 \simeq 0.1379$,
marked by the horizontal dashed line.
(b) The transverse ($y$) component of $\vec v_{\rm c.m.}$ stabilizes to 0
after the transient.
The simulation is carried out with $F_{\rm load} = 0$, $T=0$, $v_{\rm ext}
=0.05$.
The transient detail depends on several physical quantities, including the
initial configuration, the top speed $v_{\rm ext}$, temperature $T$, and
the dissipation coefficient $\eta$.
In contrast, the final value $w$ in the quantized-sliding state is
completely insensitive to these details, but only depends on the lattice
mismatch.
}
\end{figure}

A simulation will represent the steady dynamical state of the system
provided (i) that the simulation time is much longer than the relaxation
times of all quantities of interest and (ii) that it yields a sufficiently
long sampling of fluctuations to obtain accurate time averages in the
dynamical steady state.
In all our calculations we discard an initial transient, extending usually
for a comparably long time ($100$ to $1000$ time
units), related to the poor damping produced by the relatively weakly
coupled thermostat ($\eta = 0.05$).
Figure~\ref{transient:fig} illustrates a typical transient regime for the
lubricant center-mass velocity.
Over the ensuing steady running state, we evaluate the time-averages of
physical quantities.
Whenever the quantities to be averaged happen to fluctuate periodically, we
minimize systematic errors by evaluating these averages over one or several
periods.

When we run simulations with different $v_{\rm ext}$, we set the total
evolution time of each simulation $t_{\rm calc}$ by fixing the product
$t_{\rm calc} \, v_{\rm ext}$, so that in a simulation the top slider
advances by the same distance.
We take at least $v_{\rm ext}\, t_{\rm calc}= 10$ length units for each
simulation, and we also include a condition that $t_{\rm calc}$ never
decreases under $100$ time units, which is usually sufficient because when
$v_{\rm ext}$ is changed in small steps transients are shorter than the one
illustrated in Fig.~\ref{transient:fig}.
For moderate speeds $v_{\rm ext}\lesssim 1$, this choice allows the system
enough time for all initial transient stresses induced by a changed $v_{\rm
  ext}$ to relax, and for a steady sliding state to ensue.

\begin{figure}
\centerline{(a)\hfill
\includegraphics[width=80mm,angle=0,clip=]{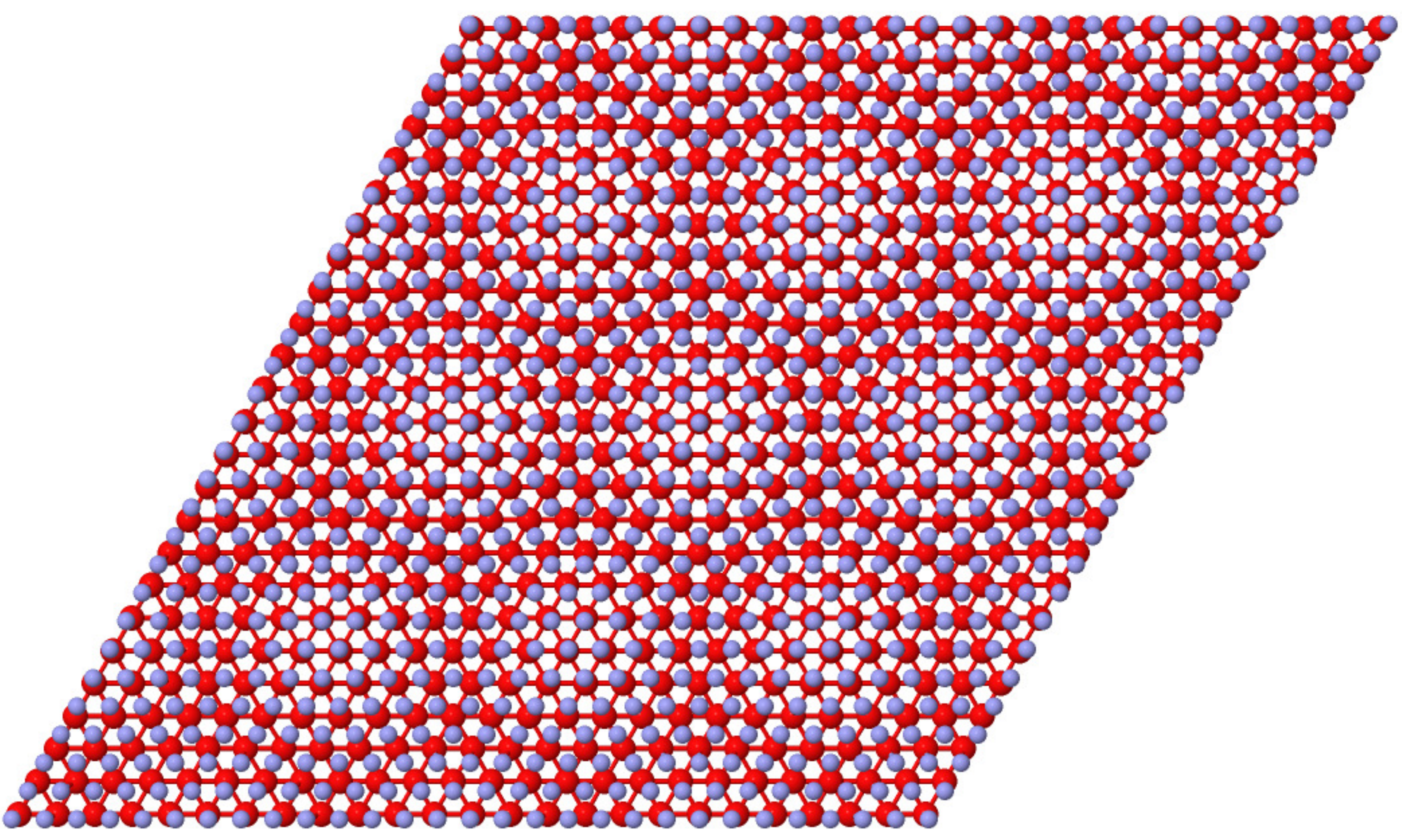}
}
\centerline{(b)\hfill
\includegraphics[width=80mm,angle=0,clip=]{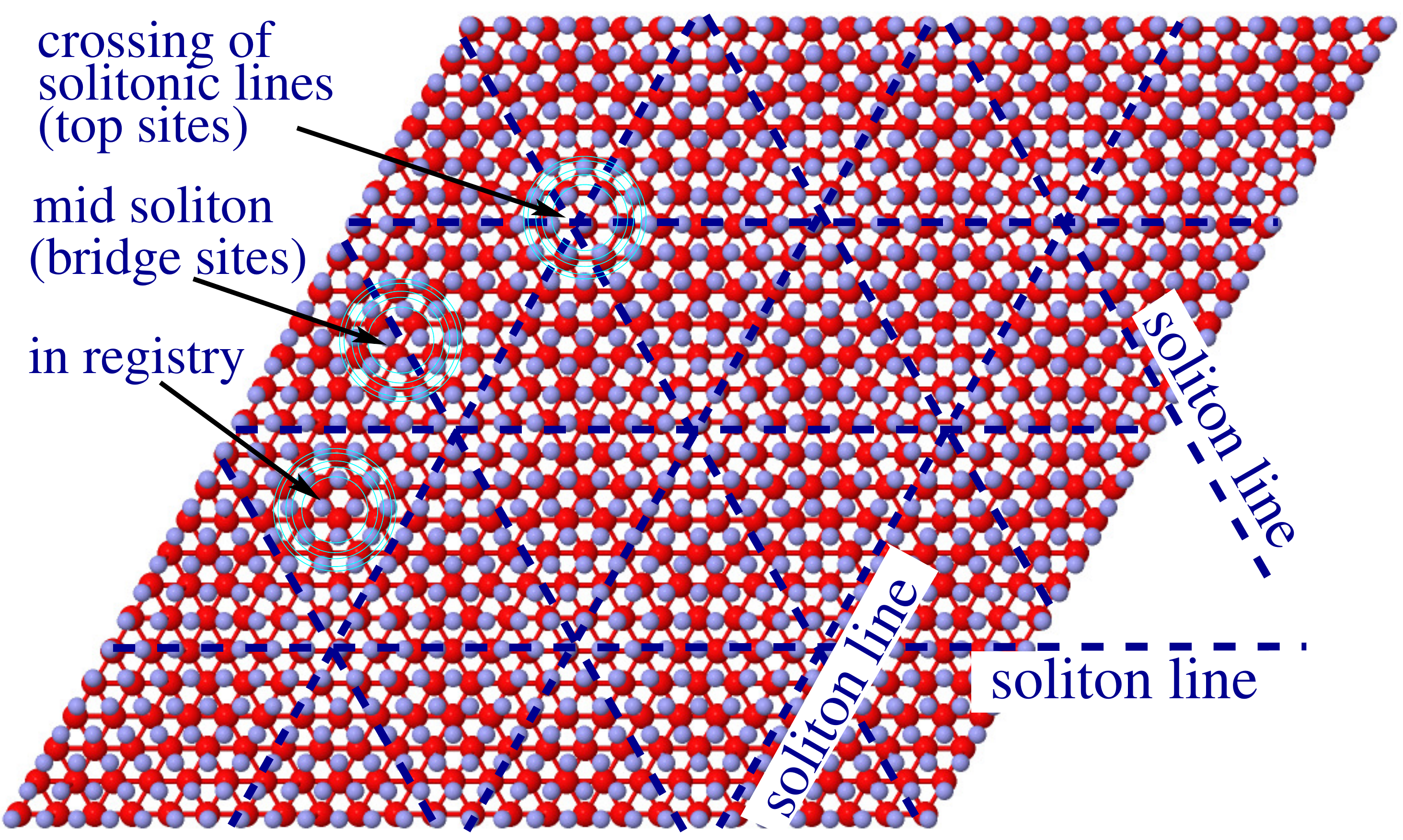}
}
\centerline{(c)\hfill
\includegraphics[width=80mm,angle=0,clip=]{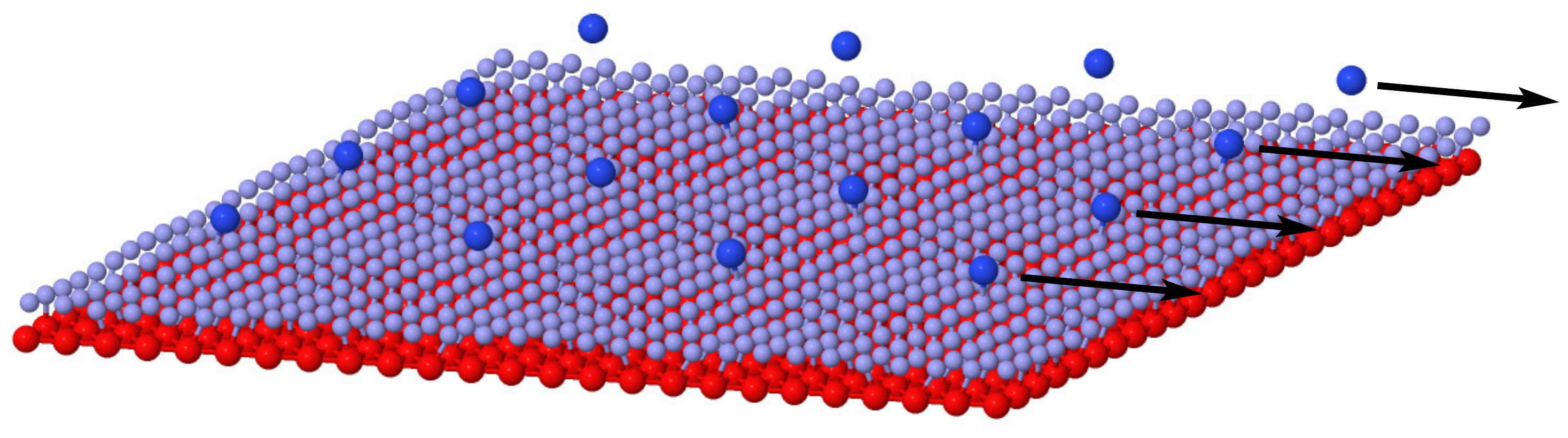}
}
\caption{\label{modelsketch} (Color online)
A snapshot of the sliding steady state of the simulation of
Fig.~\ref{transient:fig}, representing the atoms inside one PBC-repeated
supercell, in the same color convention as Fig.~\ref{snap:fig}.
The bottom slider and the lubricant are aligned with a lattice spacing
mismatch of $r_{\rm b}=a_{\rm b}/a_{\rm p}= 29/25= 1.16 $, producing a
clear soliton pattern.
(a) Top view of a lattice-mismatched configuration, where only the bottom
slider atoms (larger, red) and lubricant atoms (smaller, light blue) are
shown.
(b) Same top view as (a), with marked soliton lines and their crossings
(top-site lubricant atoms), and in-registry regions (hollow-site lubricant
atoms).
These regions form a loose triangular lattice of spacing $a_{\rm sol} =
25/4 a_{\rm b} =6.25\, a_{\rm b}$.
(c) Lateral perspective view, with the advancing top layer (dark blue
particles) spaced by $a_{\rm t}= a_{\rm sol}$ also included.
}
\end{figure}

Guided by the lesson learnt in earlier 1D models \cite{Vanossi06,
  Vanossi07PRL} -- solitons formed in the lubricant by one slider are
docked and dragged by the other slider -- we adopt a geometry of
near-commensuration of the lubricant spacing to that of the bottom slider,
with $r_{\rm b}$ not far from unity, and $r_{\rm t}$ far away from unity
and closer instead to commensurate with the soliton lattice.
Figure~\ref{modelsketch} displays a configuration of this kind, which we
adopt as a prototype in the present paper.

\subsection{Quantized lubricant sliding}\label{qss:sec}

Simulations show that in most cases the lubricant slides relative to the
bottom substrate with a relative mean lubricant velocity component in the
driving direction $w=v_{{\rm c.m.}\,x}/v_{\rm ext}$ giving rise to a plateau
$w= w_{\rm quant}$ which is essentially parameter-independent, that is
``quantized'' as in the the more idealized models studied in the past.
We ran several batches of MD simulations to characterize the properties and
boundaries of this plateau of quantized-velocity dynamics in the 3D model.
To evaluate the dragging of solitons and the ensuing velocity-quantization
phenomenon in 3D, for each lubricant layer we need to compute the mean flux
$\bar\Phi_{\rm p}$ of lubricant particles crossing a line of length $L_y$
transverse to the pulling direction.
By dividing $\bar\Phi_{\rm p}$ by a hypothetical flux $\bar\Phi_{\rm
  p}^{v_{\rm ext}}$ of lubricant particles all moving across the $L_y$ line
at speed $v_{\rm ext}$, we obtain
\begin{equation}\label{w_flux}
w \equiv \frac{ v_{{\rm c.m.}\,x}}{v_{\rm ext}}
\equiv
\frac{\bar\Phi_{\rm p}}{\bar\Phi_{\rm p}^{v_{\rm ext}}}
\,.
\end{equation}

\begin{figure}
\centerline{
\includegraphics[width=70mm,angle=0,clip=]{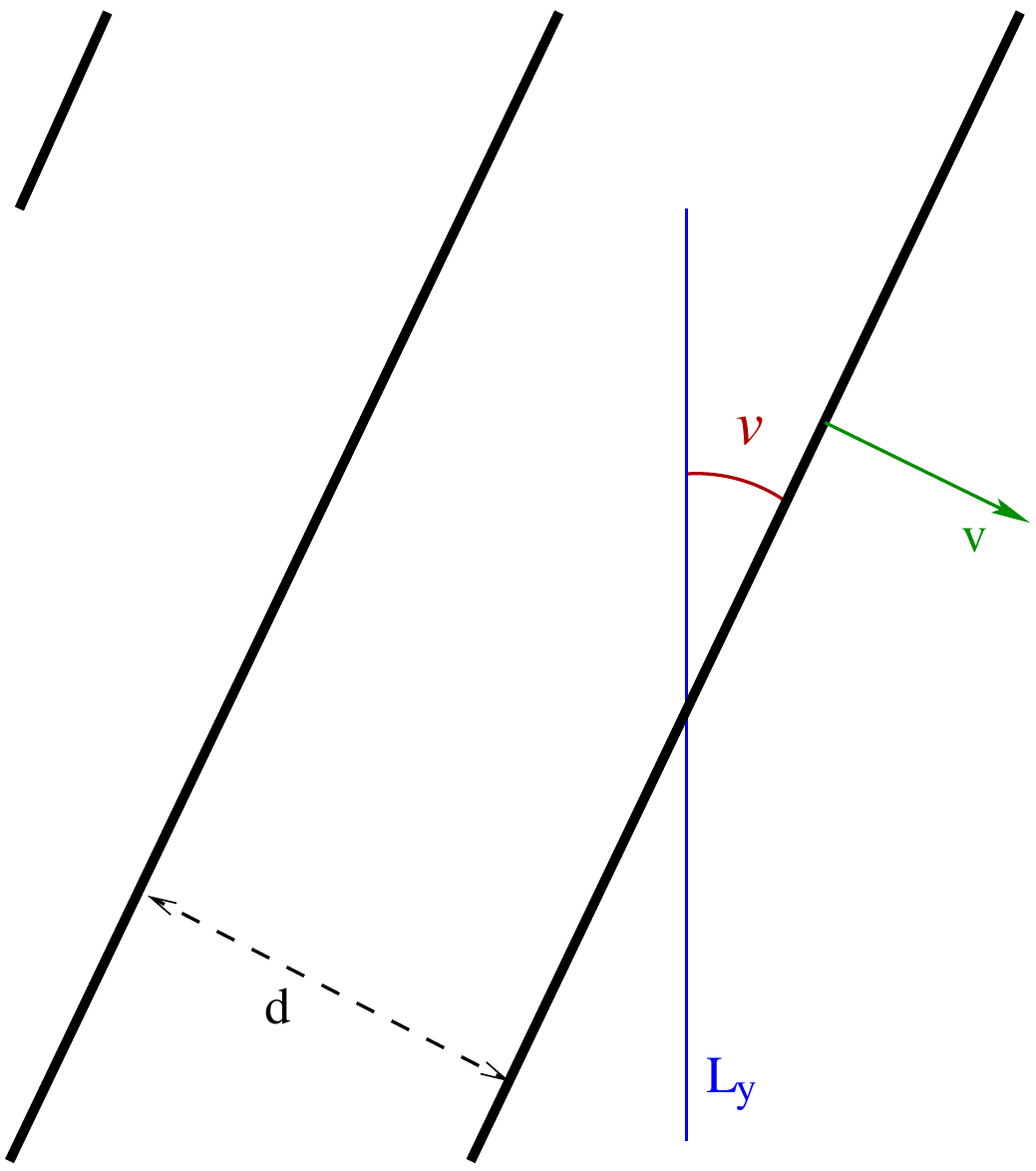}
}
\caption{\label{flux1:fig} (Color online)
A train of lines, spaced by a distance $d$, moves perpendicular to their
direction at speed $v$ and crosses a segment of length $L_y $ at an angle
$\nu$.
This construction allows us to evaluate the length of line crossing the
segment per unit time, Eq.~\eqref{eq_meanflux}.
}
\end{figure}

Firstly, we evaluate the length $\delta_{\rm sol}$ of a single soliton line
that crosses our reference line $L_y$ in a time $\tau$, while advancing
perpendicularly to its own elongation
\begin{equation}\label{eq_fluxsoliton}
\delta_{\rm sol} = v \, \tau \, \frac{\cos \nu}{\sin \nu}
\end{equation} 
where $\nu$ is the angle formed by the soliton line with the $L_y$
direction, also equaling the angle that the advancement direction makes
with the pulling direction, see Fig.~\ref{flux1:fig}.
We then evaluate the mean length of soliton lines crossing $L_y$ in a unit
time for a train of parallel soliton lines separated by a mutual distance
$d$:
\begin{equation}\label{eq_meanflux}
\bar V =
 \frac{\delta_{\rm sol}}{\tau} \, \frac{L_y \sin \nu / v} {d/v}
=
 v\, \frac{\cos \nu}{\sin \nu} \, \frac{L_y \sin \nu}{d}
=
 \frac{L_y}{d}\, v\, \cos\nu
\,,
\end{equation}
where $d/v$ represents the time between two successive solitons starting to
cross $L_y$, and $L_y \sin \nu / v$ the time it takes for one such crossing
to occur.

\begin{figure}
\centerline{
\includegraphics[width=75mm,angle=0,clip=]{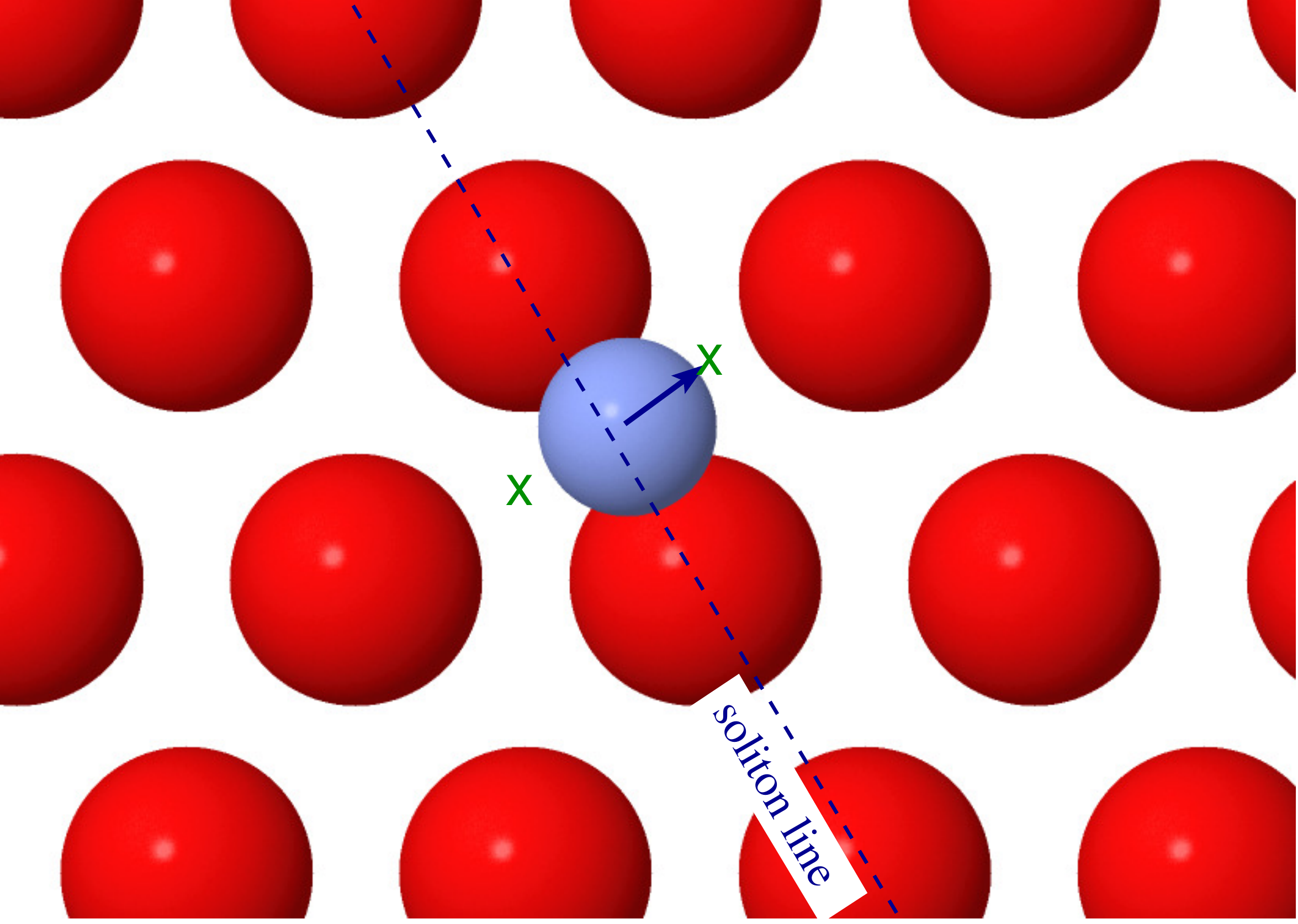}
}
\caption{\label{hollowBRIDGEhollow:fig} (Color online)
A typical bridge atom along a soliton line (dashed line) moves from one
hollow site to the next (green crosses), thus advancing perpendicularly to
the soliton line.
}
\end{figure}

We first apply this general result to the case of a soliton pattern formed
by a lattice-spacing mismatch between two aligned triangular lattices.
In terms of the spacing $a_{\rm sol}$ of the lattice of soliton-crossing
areas, see Fig.~\ref{modelsketch}b, successive soliton lines are separated
by $d=\frac{\sqrt{3}}{2}\, a_{\rm sol}$.
The soliton spacing in the aligned case is given \cite{Braunbook} by the 1D
geometric mismatch condition
\begin{equation}\label{1D_soliton_density}
a_{\rm sol}^{-1} = a_{\rm p}^{-1}-a_{\rm b}^{-1}
\,.
\end{equation}
A soliton line can only advance perpendicularly to itself, because the
soliton-forming atoms stand locally at bridge sites relative to the bottom
surface: Each atom is forced to cross the saddle-point energy barrier
between highly-coordinated hollow sites moving in the energetically most
favorable direction, which is perpendicular to the soliton line, see
Fig.~\ref{hollowBRIDGEhollow:fig}.
The soliton intersections are dragged forward by the top layer moving at
speed $v_{\rm ext}$.

\begin{figure}
\centerline{(a)
\includegraphics[width=80mm,angle=0,clip=]{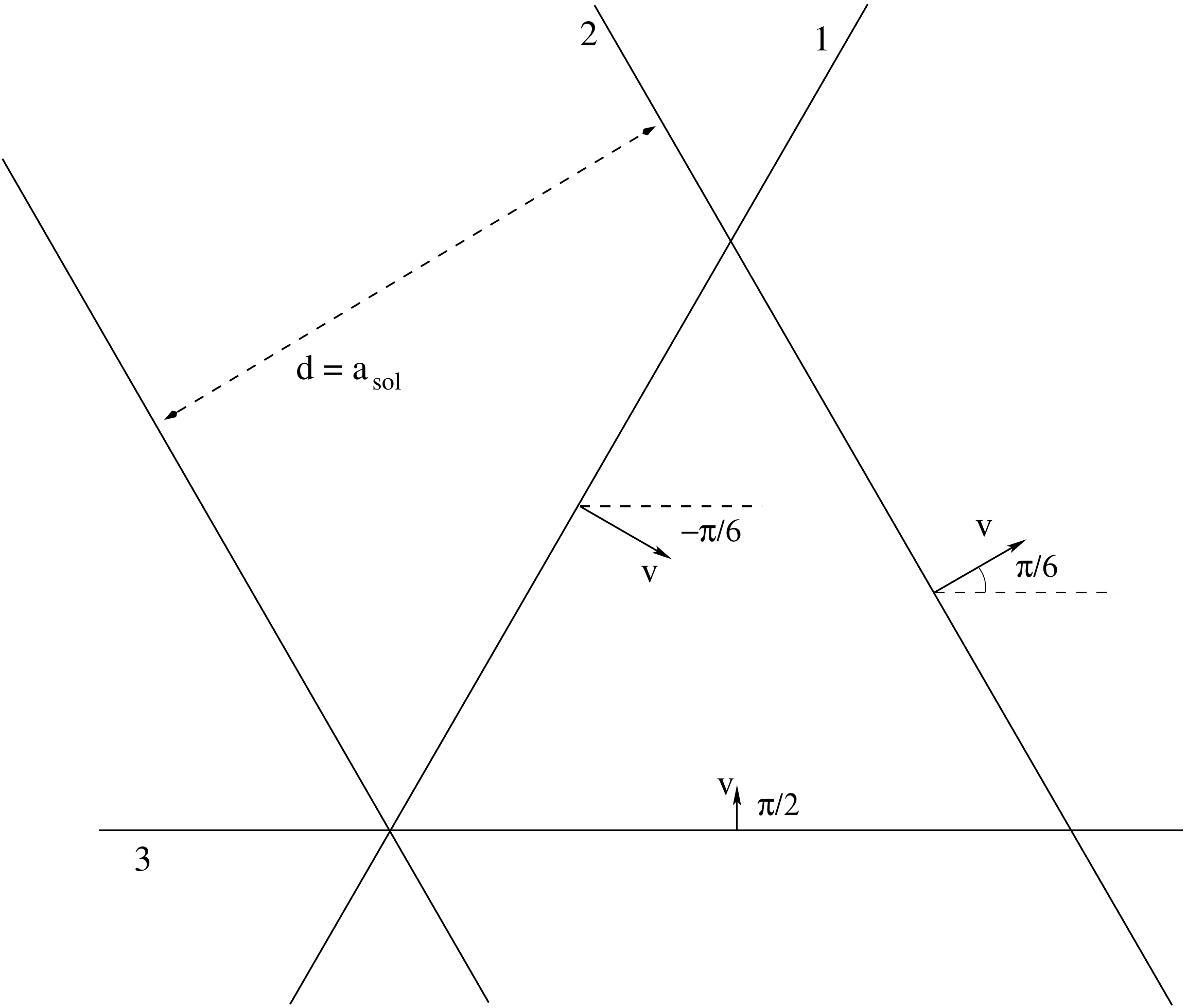}
}\centerline{(b)
\includegraphics[width=80mm,angle=0,clip=]{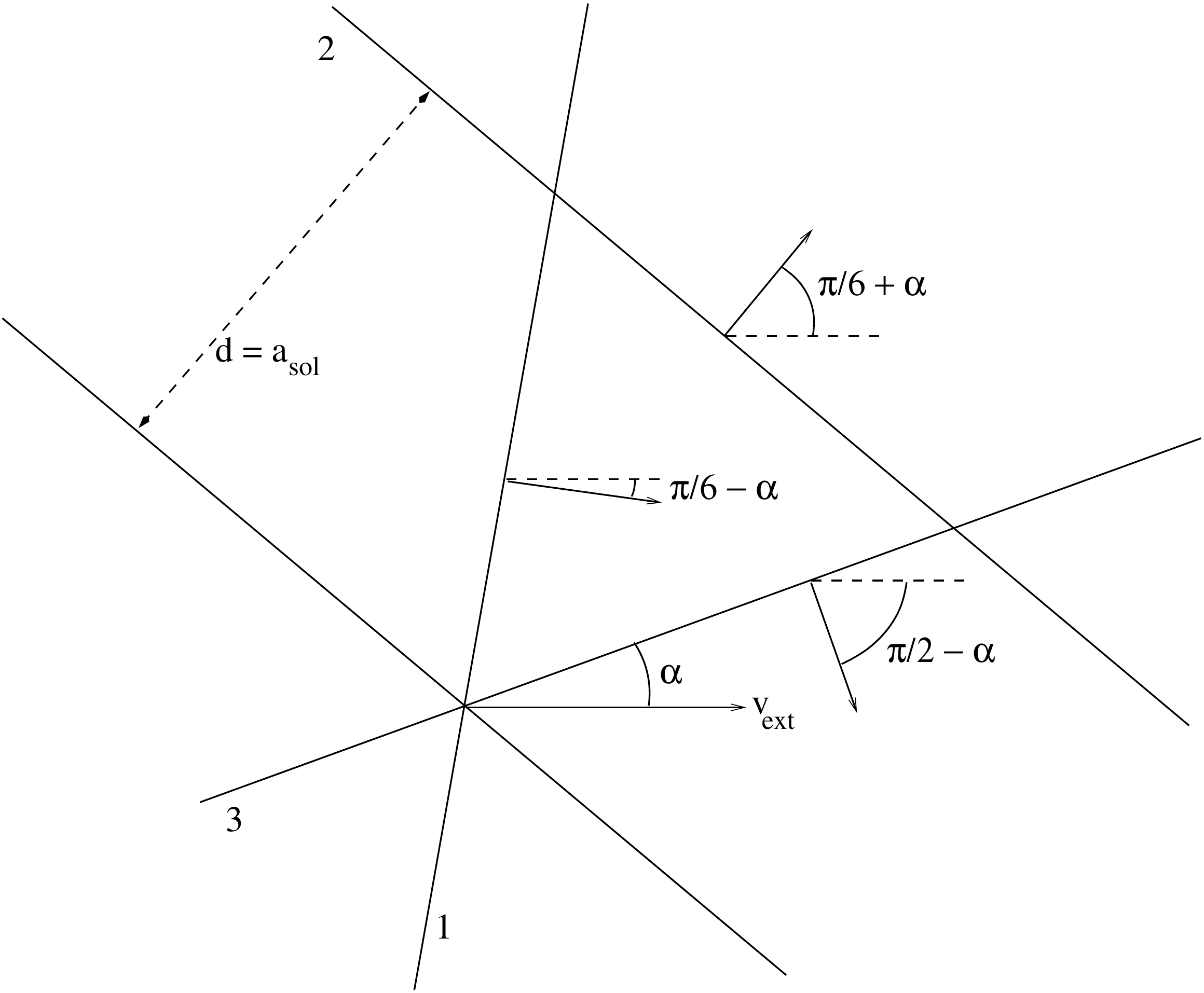}
}
\caption{\label{flux2:fig}
The geometry of a triangular lattice of soliton lines moving perpendicular
to their direction at a speed $v$.
(a) The case where the lines of type $3$ are parallel to the dragging
direction $\hat x$.
(b) The case characterized by an overall rotation by $\alpha$ relative to
the dragging direction $\hat x$.
}
\end{figure}

This advancement is realized when each one of the soliton lines advances
perpendicularly to itself at a speed $v=v_{\rm ext} \,\cos\nu$, namely a
speed scaled by the angle that each line forms with the dragging direction.
In the simplest case of unrotated lattices, see Fig.~\ref{flux2:fig}a, of
the three soliton families, the one labeled $3$ is horizontal, namely
perpendicular to $L_y$, thus it does not contribute to the rightward
sliding of the lubricant ($\nu=\pi/2$, thus $v=0$).
The two other families of solitons, labeled $1$ and $2$, both contribute a
speed reduced by a factor $\cos \nu = \cos(\pi/6)$.

Using Eq.~\eqref{eq_meanflux}, we evaluate the total speed of soliton lines
crossing $L_y$ in this unrotated case, obtaining
\begin{eqnarray}\label{eq_flux1}
\bar V_1 &=&
\frac{L_y v_{\rm ext}\cos\frac{\pi}{6}\cos\frac{\pi}{6}}
{a_{\rm sol}\sqrt{3}/2} =
\frac{2 L_y v_{\rm ext}}{\sqrt{3} a_{\rm sol}} \cos^2\frac{\pi}{6}\\
\label{eq_flux2}
\bar V_2 &=& \bar V_1\\
\label{eq_fluxtot}
V_{\rm tot} &=& \bar V_1+\bar V_2=
\frac{\sqrt{3}L_y v_{\rm ext}}{a_{\rm sol}}
\,,
\end{eqnarray}
where $V_{\rm tot}$ includes contributions from all advancing soliton lines.

As a next step, we evaluate the flux of mobile particles associated with the
advancing soliton lines.
Recalling Fig.~\ref{modelsketch}b, we observe that: (i) in-registry
particles in between solitons do not contribute to sliding, as they are
trapped in individual minima of the corrugation potential; (ii) a soliton
line represents a single line of extra particles; (iii) as the soliton
lines are parallel to the crystal principal directions, the line density of
such extra particles along a soliton line is simply the reciprocal lattice
spacing of the lubricant, $a_{\rm p}^{-1}$; (iv) one half of each soliton
is composed of particles in the region in between soliton-crossing areas
(bridge overlayer sites), which belong uniquely to that soliton, while the
other half particles, those in the soliton crossing region (top sites), are
shared by three solitons, thus the effective mean line density of mobile
soliton particles is $\frac 12 \times (1 + \frac 13) a_{\rm p}^{-1} = \frac
23 a_{\rm p}^{-1}$.
By multiplying this atomic linear density by $V_{\rm tot}$, we obtain the
total flux of particles crossing $L_y$ per unit time, due to soliton
advancement
\begin{equation}\label{eq_fluxparticles}
\bar\Phi_{\rm p} =
\frac{2 L_y v_{\rm ext}}{\sqrt{3}\,  a_{\rm p} a_{\rm sol}}
\,.
\end{equation}
We can now evaluate the dimensionless ratio of Eq.~\eqref{w_flux}:
\begin{equation}
w =
\frac{\bar\Phi_{\rm p}}{\bar\Phi_{\rm p}^{v_{\rm ext}}}
=
\frac{
\frac{2 L_y v_{\rm ext}}{\sqrt{3}\,  a_{\rm p} a_{\rm sol}}
}{\frac{L_y v_{\rm ext}}{\sqrt{3}\,  a_{\rm p}^{2}/2}}
= \frac{a_{\rm p}}{a_{\rm sol}}
\,.
\end{equation}
This expression is independent not just of $L_y$ but also of $v_{\rm ext}$,
and it is a purely geometric function of the crystal lattice spacings which
we can make explicit using Eq.~\eqref{1D_soliton_density} for $a_{\rm
  sol}$:
\begin{equation}\label{w_quant}
w=a_{\rm p} \left(\frac{1}{a_{\rm p}}-\frac{1}{a_{\rm b}}\right)
=
1-\frac{a_{\rm p}}{a_{\rm b}}
=
1-\frac{1}{r_{\rm b}}
\equiv w_{\rm quant}
\,.
\end{equation}
This formula coincides with the 1D result \cite{Vanossi07PRL} and matches
the outcome of simulations as discussed in the next section.

In the case of a rigid overall rotation by a common angle $\alpha$, we
apply the same theory, but we need to re-evaluate the speed of soliton
lines crossing a line $L_y$ directed perpendicularly to the dragging
direction $\hat x$.
Using Eq.~(\ref{eq_meanflux}), we evaluate the crossing speed of the three
families of parallel soliton lines shown in Fig.~\ref{flux2:fig}b:
\begin{equation}\label{eq_fluxrigid}
\begin{split}
\bar V_1 &= \frac{L_y\,v_{\rm ext} }{a_{\rm sol}\frac{\sqrt{3}}{2}}
\cos^2\left( \frac{\pi}{6}-\alpha \right)\\
\,
\bar V_2 &= \frac{L_y\,v_{\rm ext} }{a_{\rm sol}\frac{\sqrt{3}}{2}}
\cos^2\left( \frac{\pi}{6}+\alpha \right)\\
\,
\bar V_3 &= \frac{L_y\,v_{\rm ext}}{a_{\rm sol}\frac{\sqrt{3}}{2}}
\cos^2\left( \frac{\pi}{2}-\alpha \right)
\,.
\end{split}
\end{equation}
By summing these three contributions, we obtain
\begin{eqnarray}\label{eq_fluxrigidtotal}
V_{\rm tot} &=& \bar V_1+ \bar V_2+ \bar V_3  \\\nonumber
&=& \frac{L_y\, v_{\rm ext}}{a_{\rm sol}\frac{\sqrt 3}{2}} \left[
   \frac{3}{2}   \cos^{2} \alpha +\frac{3}{2} \sin^{2} \alpha \right] =
\frac{\sqrt{3} L_y v_{\rm ext}}{a_{\rm sol}}
\,,
\end{eqnarray}
which coincides with the unrotated result, Eq.~\eqref{eq_fluxtot}.
As also the particle density along the soliton lines is the same, we
obtain the same particle flux, as given by Eq.~\eqref{w_quant}.
We conclude that an overall rotation produces no change in the quantized
sliding state, consistently with the fundamental isotropy of the triangular
soliton net.

\subsection{Dragging solitons: forward lubricant motion}

Zero-temperature MD simulations confirm the phenomenon of perfect velocity
quantization in both the unrotated and the rigidly rotated case.
As an example, the unrotated single-layer case of the model of
Fig.~\ref{modelsketch} is characterized by $a_{\rm b} = 25/29$, thus
$r_{\rm b} = 29/25$ and $a_{\rm sol} = 29/4$: this indicates that we cross
$4$ soliton lines every $29$ lubricant particles in each 2D-crystal
high-symmetry direction.
The choice $r_{\rm t} = a_{\rm sol}$ guarantees that for each one of these
soliton lines, a line of top atoms is there to grab it.
Figure~\ref{transient:fig}a compares the instantaneous center-mass
lubricant speed to the predicted quantized value of Eq.~\eqref{w_quant}:
after the initial transient, the resulting $v_{{\rm c.m.}\,x}$ makes a tiny
oscillation around $w_{\rm quant} = 4/29 \simeq 0.1379$.
This value of $r_{\rm b} $ is not to be considered in any way special: we
find perfect quantized sliding for many other values of $r_{\rm b}$.

\begin{figure}
\centerline{
\includegraphics[width=80mm,angle=0,clip=]{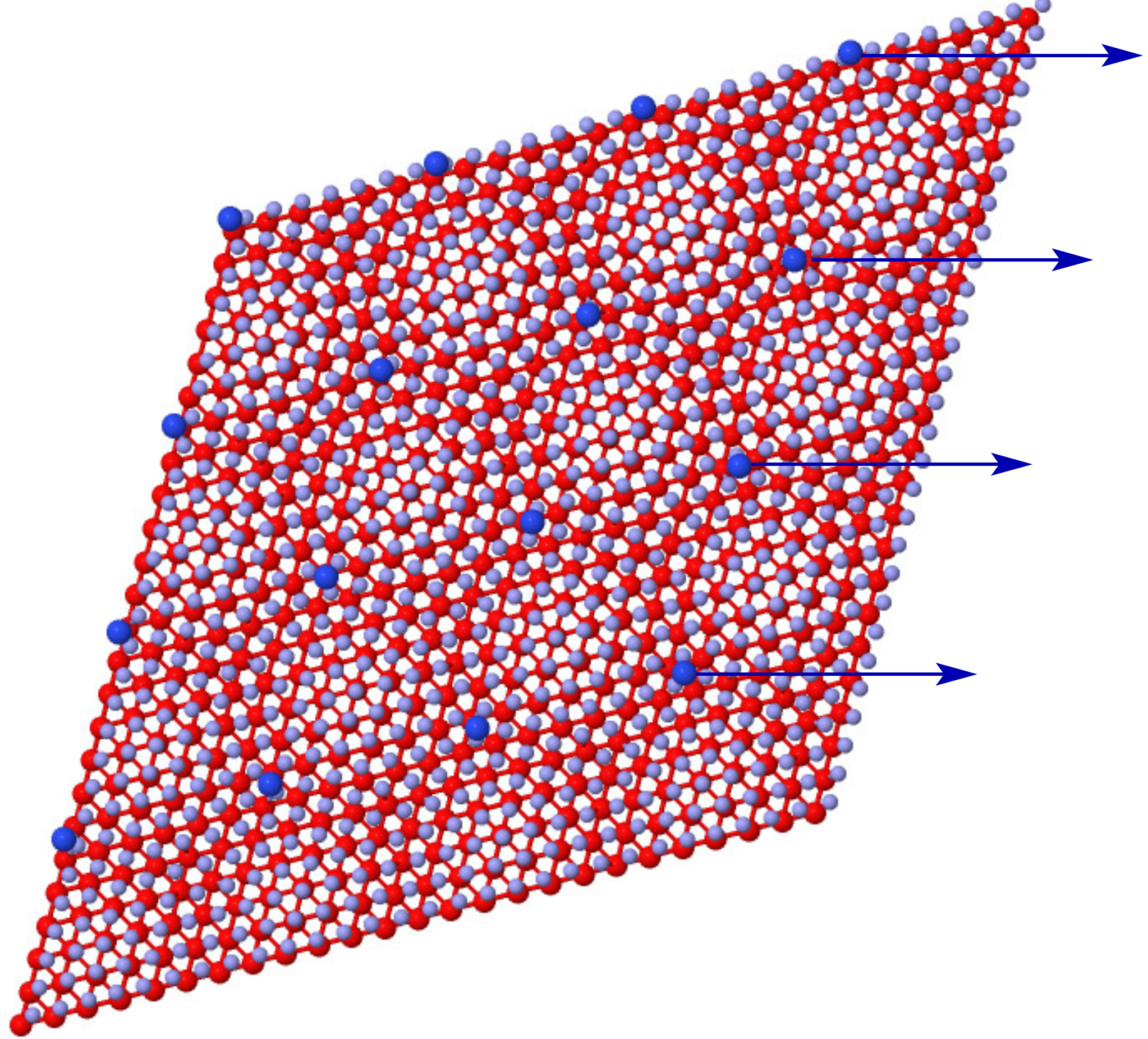}
}
\caption{\label{rotationAA:fig} (Color online)
Top view of the same model as in Fig.~\ref{modelsketch}, but rotated
rigidly by an angle $\alpha=\pi/12$ with respect to the original
orientation.
The top layer is still driven in the same horizontal direction $\hat x$
highlighted by arrows.
}
\end{figure}

\begin{figure}
\centerline{
\includegraphics[width=85mm,angle=0,clip=]{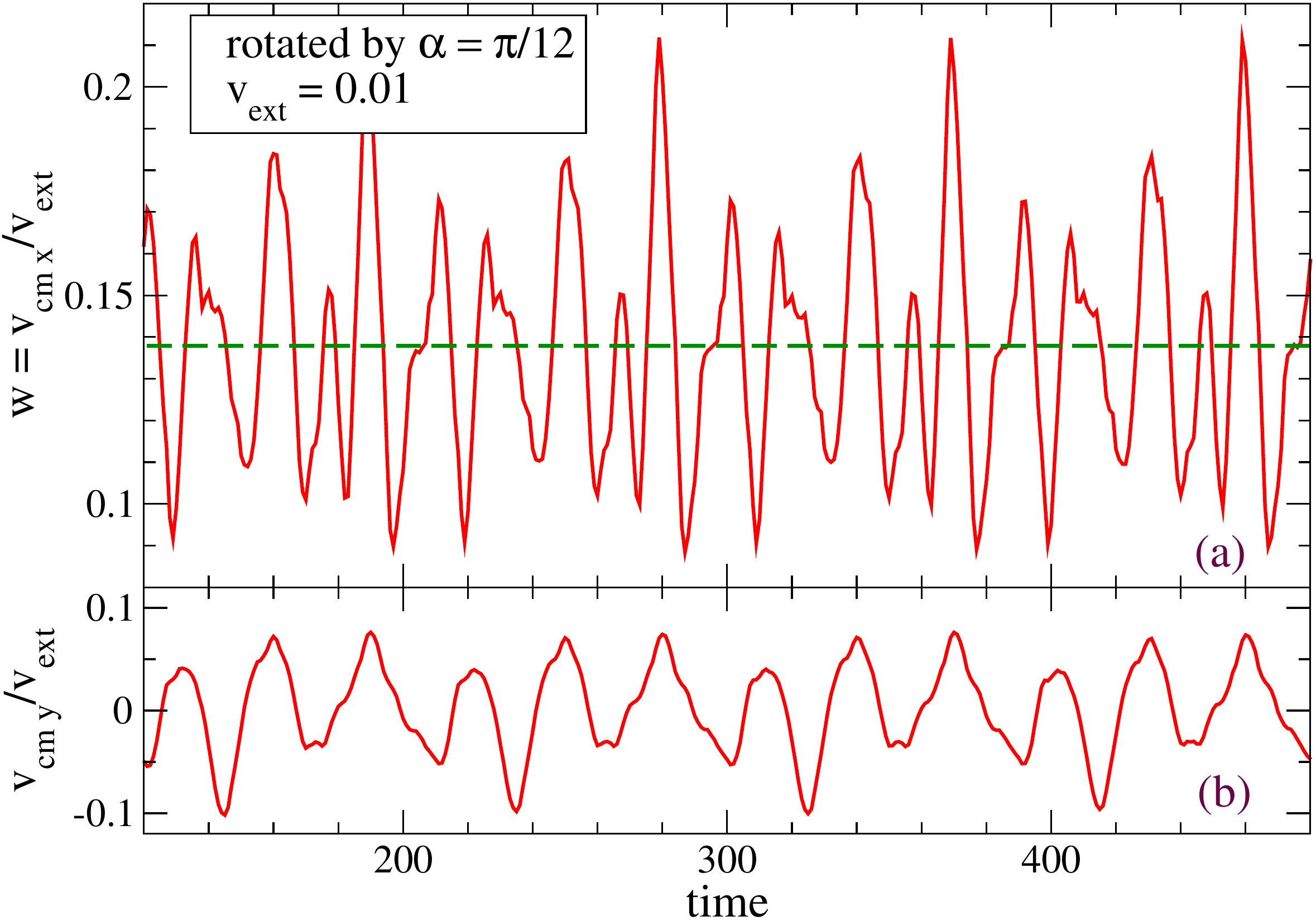}
}
\caption{\label{rotationBB:fig} (Color online)
(a) $\hat x$ and (b) $\hat y$ components of the lubricant center-mass
  velocity divided by the driving speed, $v_{\rm ext}=0.01$, as a function
  of time for the rotated model of Fig.~\ref{rotationAA:fig}.
The time-averaged center-mass $\hat x$ component coincides with the one
obtained for the unrotated model, Fig.~\ref{transient:fig}, and matches the
quantized formula \eqref{w_quant}, dashed line.
The amplitude and period of the fluctuations of $w$ around $w_{\rm quant}$
are both substantially larger than in the unrotated case,
Fig.~\ref{transient:fig}.
The average $v_{{\rm c.m.}\,y}$ is consistent with the lubricant moving at an
average angle of $1.1^\circ$ with the $\hat x$ driving direction.
}
\end{figure}

Likewise, by rotating rigidly the model of Fig.~\ref{modelsketch} e.g.\ by
an angle $\alpha=\pi/12$, we obtain the geometry sketched in
Fig.~\ref{rotationAA:fig}.
When we pull the top slider along the same horizontal direction $\hat x$,
we obtain the time evolution of the center of mass displayed in
Fig.~\ref{rotationBB:fig}.
Again, this velocity oscillates periodically around to the same relative
value $w_{\rm quant}$, as predicted by Eq.~\eqref{w_quant}, but with a
different oscillation pattern.
The longer period and larger oscillation amplitude are related to the
necessity of a coordination of the forward motion with a transverse motion,
induced by the tendency of the lubricant to follow the grooves of the
bottom substrate, and detected as a nonzero average of the transverse
velocity component, Fig.~\ref{rotationBB:fig}b.
We explored different rotation angles $\alpha$.
For comparatively small $|\alpha|\lesssim 15^\circ$ and small $v_{\rm
  ext}$, a similar transverse motion establishes, characterized by periodic
oscillations of the center-mass speed; for larger (nontrivial) $\alpha$ and
for intermediate driving speed little or no substrate channeling nor net
transverse motion arises, with the result that the center-mass motion is
apparently non periodic (or of extremely long period).
This is due to the advancing lubricant layer exploring the bottom-layer
corrugation in an ever renewed mutual configuration.
Importantly, in all tested cases, $w$ fluctuates (periodically or
nonperiodically) around $w_{\rm quant}$, as long as $v_{\rm ext}$ is not
too large.

\subsection{Dragging antisolitons: backward lubricant motion}
\label{antisolitons:sec}

\begin{figure}
\centering
\includegraphics[width=80mm,angle=0,clip=]{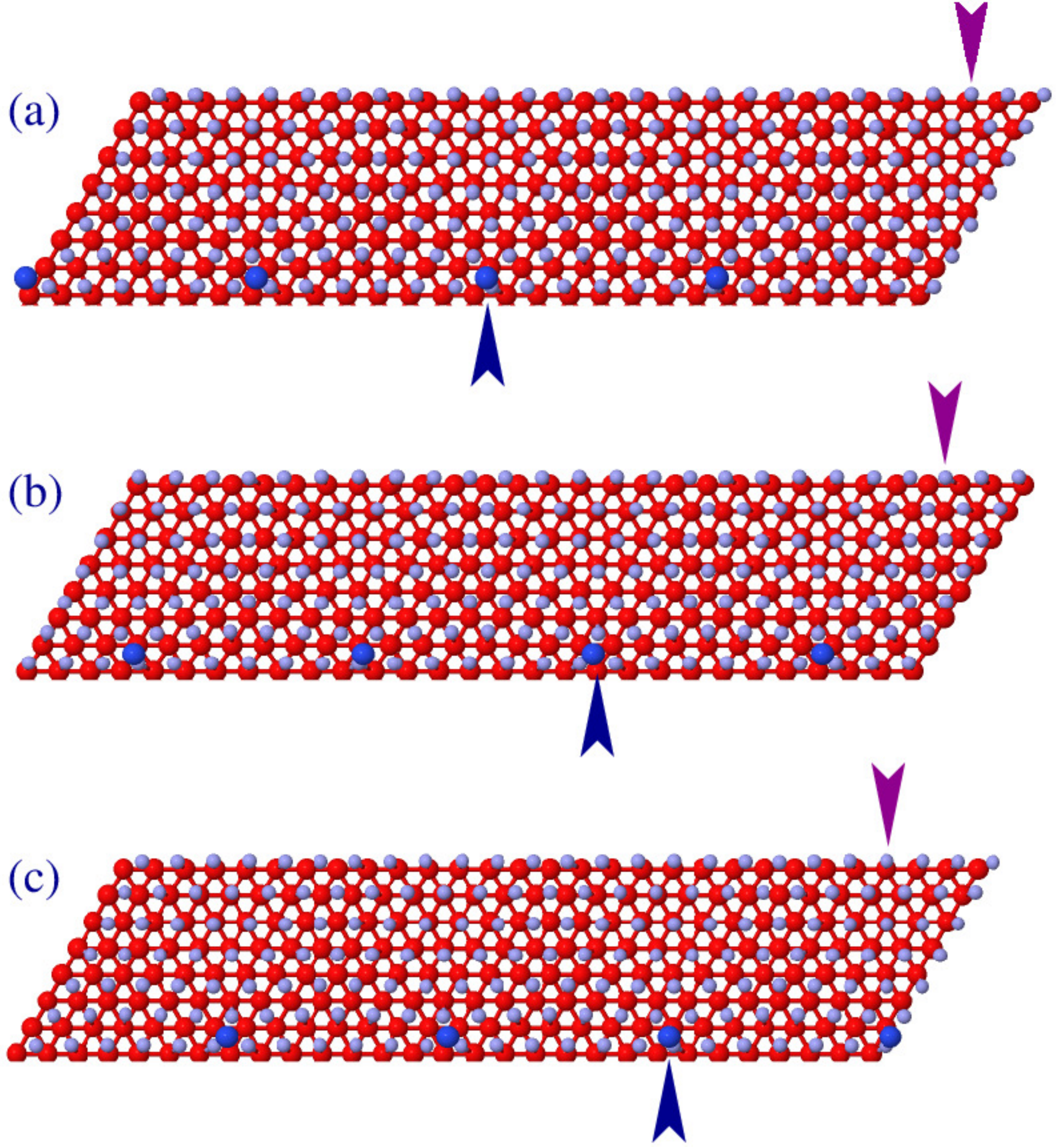}
\caption{\label{antisolitons:fig} (Color online)
A portion of three successive snapshots of the steady state of an
underdense lubricant layer ($r_{\rm b}= 25/29$) forming a Moir\'e pattern,
with $\Theta=1$ antisoliton every top-layer line (same atomic symbol
convention as in previous figures).
Arrows track two atoms to help visualizing the leftward motion of the
lubricant induced by a rightward motion of the top layer.
Between snapshots (b) and (c) the leftmost line of lubricant particles has
been remapped back inside the cell at its right side by the PBC.
}
\end{figure}

A peculiar reversed lubricant dragging occurs when the lubricant is less
dense than the bottom layer, i.e.\ $r_{\rm b} < 1$.
Lines of dilation (antisolitons) are separated by in-register regions, as
shown in Fig.~\ref{antisolitons:fig}.
These antisoliton lines are soft defects with an enhanced mobility
similar to that of the solitons of overdense layers: They can therefore be
dragged rightward by the advancing top slider.
Since these rightward traveling antisoliton lines are basically lines of
missing atoms, or vacancies, the involved atoms, and thus the overall
lubricant center of mass, move leftward, opposite to the driving $v_{\rm
  ext}$.
As illustrated by the sequence of Fig.~\ref{antisolitons:fig}, a net
backward lubricant motion ($v_{\rm c.m.}<0$) is indeed observed.
This result is perfectly accounted for by Eq.~\eqref{1D_soliton_density},
which yields negative $a_{\rm sol}$, and by Eq.~\eqref{w_quant}, which
yields negative $w_{\rm quant}$.

\begin{figure}
\centerline{
\includegraphics[width=85mm,angle=0,clip=]{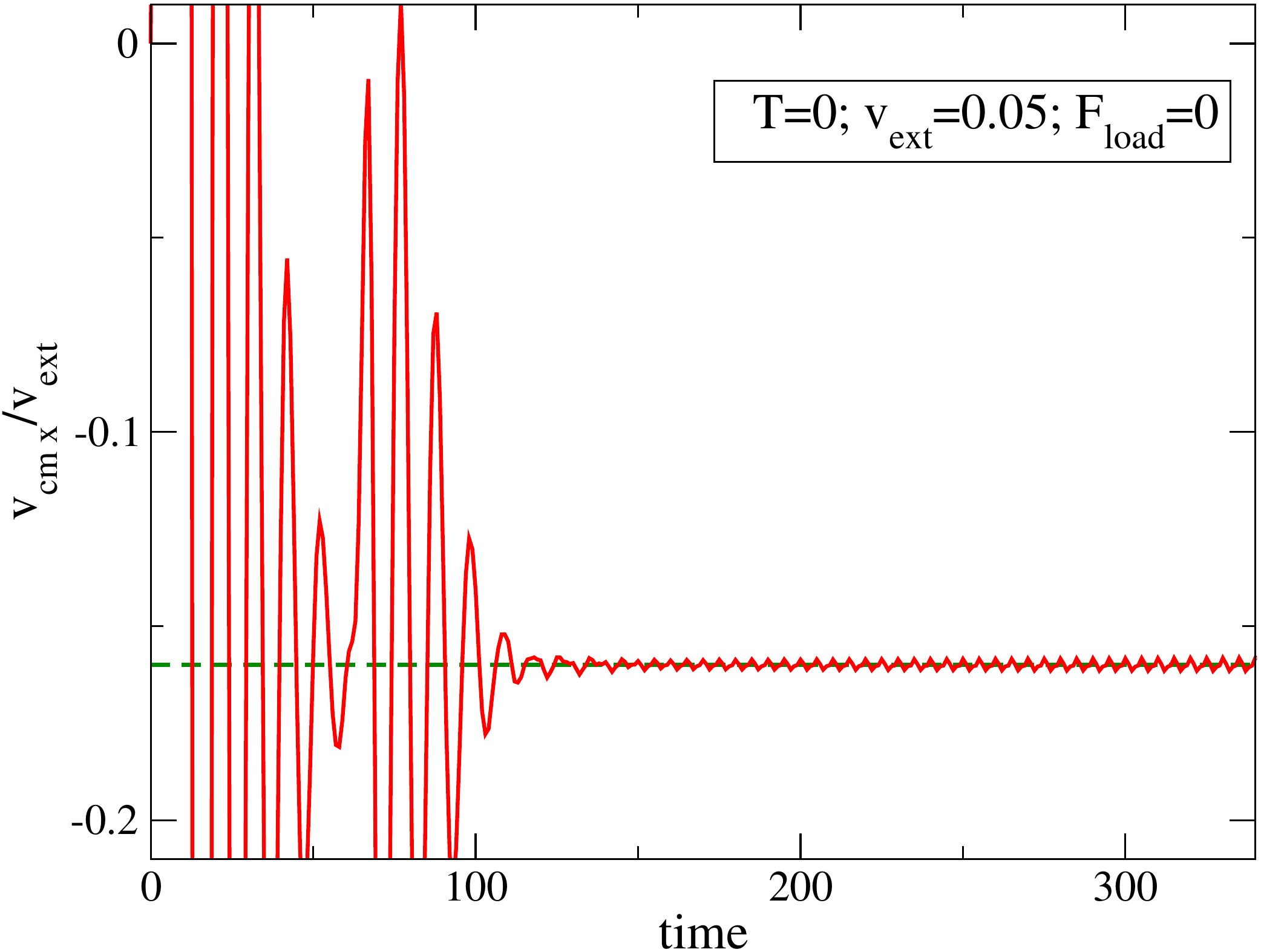}
}
\caption{\label{antisolitons_time:fig} (Color online)
The normalized average lubricant velocity, $w=v_{{\rm c.m.}\,x}/v_{\rm ext}$
as a function of time for the antisoliton geometry represented in
Fig.~\ref{antisolitons:fig}, with $a_{\rm t} =29/4=7.25$, $a_{\rm p}
=29/25=1.16$, $a_{\rm b}= 1$ for the top, lubricant and bottom layers
corresponding to $r_{\rm b}=25/29=0.86$.
After an initial transient, $w$ starts to fluctuate around the negative
value predicted by Eq.~\eqref{w_quant}: $w_{\rm quant}= -4/25 = -0.160 $,
marked by the horizontal dashed line.
The simulation is carried out for $N_{\rm layer}=1$, $v_{\rm ext}=0.05$,
$F_{\rm load} = 0$, and $T= 0$.
}
\end{figure}

The detailed example of this antisoliton case shown in
Fig.~\ref{antisolitons:fig} has $r_{\rm b} = 25/29$, so that the mismatch
generates $4$ antisoliton lines every $25$ lubricant lattice spacings.
We consider a top slider with $r_{\rm t}=25/4$, to full commensuration with
the antisoliton lattice, i.e.\ $\Theta =1$.
As reported in Fig~\ref{antisolitons_time:fig}, after the usual transient,
simulations do show a net negative lubricant velocity oscillating around
$v_{{\rm c.m.}\,x} / v_{\rm ext} =-0.16$, matching the predicted $w_{\rm
  quant} = -4/25$.

\subsection{Ar on graphite and other possible experimental realizations}

The experimentally accessible configuration of Ar layers interposed as a
lubricant in between a graphite substrate and a suitably nano-patterned top
layer is a promising system where an antisoliton dragging can occur.
The Ar monolayer is well know to be incommensurate to the graphite
substrate \cite{Bruch07}, thus its soliton pattern is likely mobile.

\begin{figure}
\centering
(a)\includegraphics[width=80mm,angle=0,clip=]{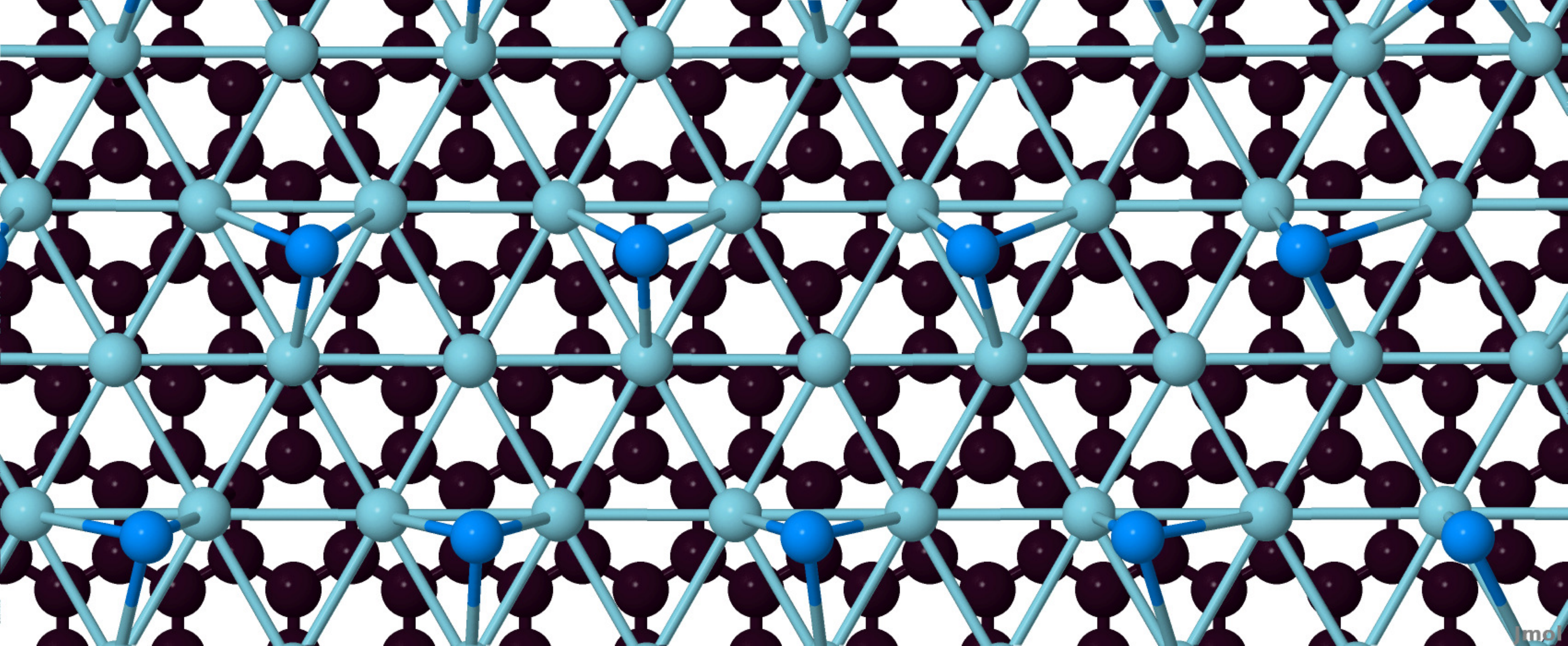}
(b)\includegraphics[width=80mm,angle=0,clip=]{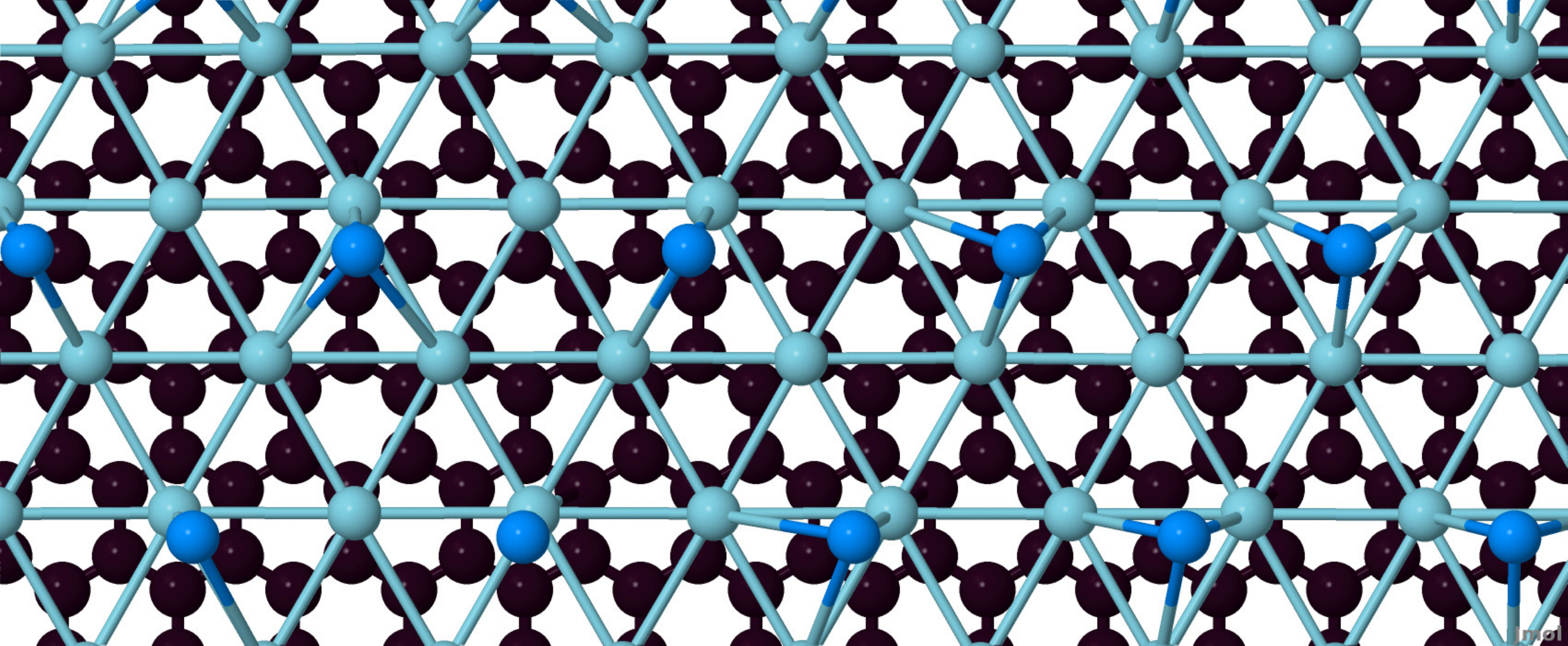}
(c)\includegraphics[width=80mm,angle=0,clip=]{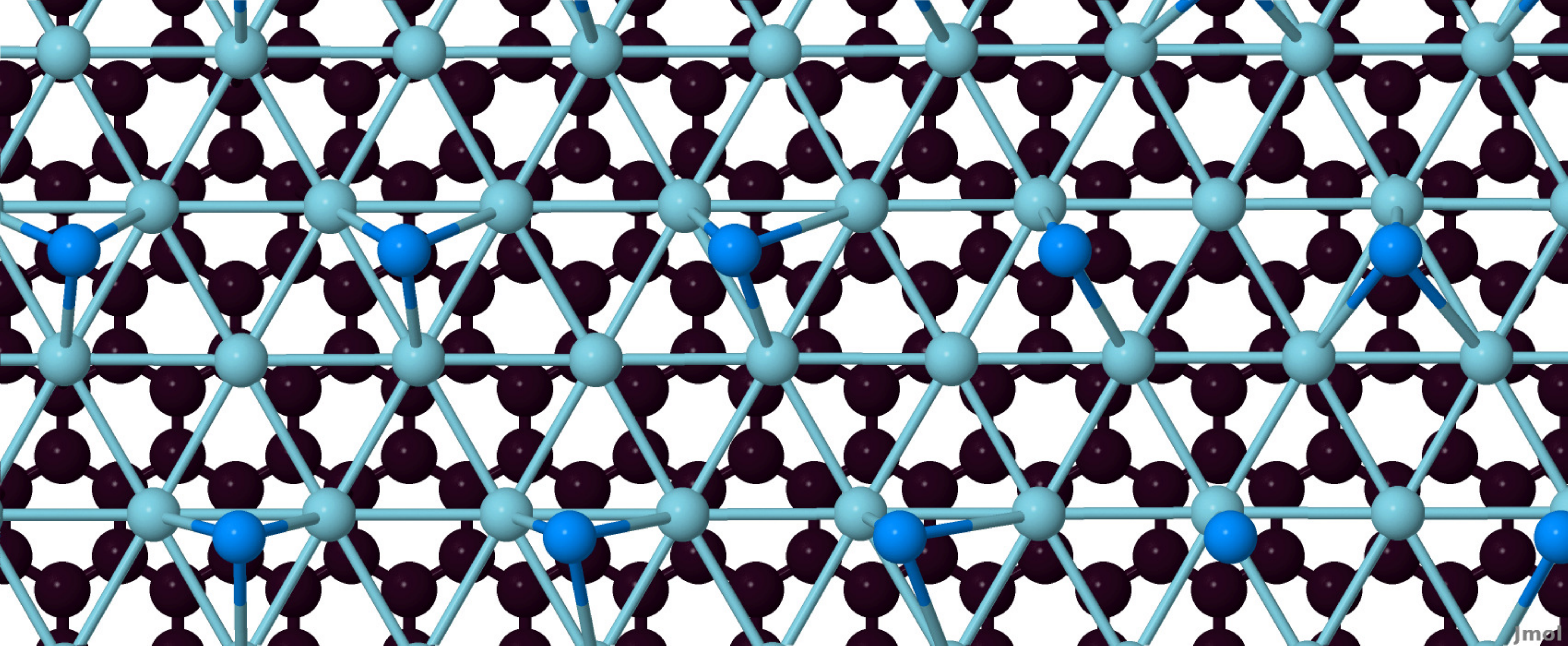}
(d)\includegraphics[width=80mm,angle=0,clip=]{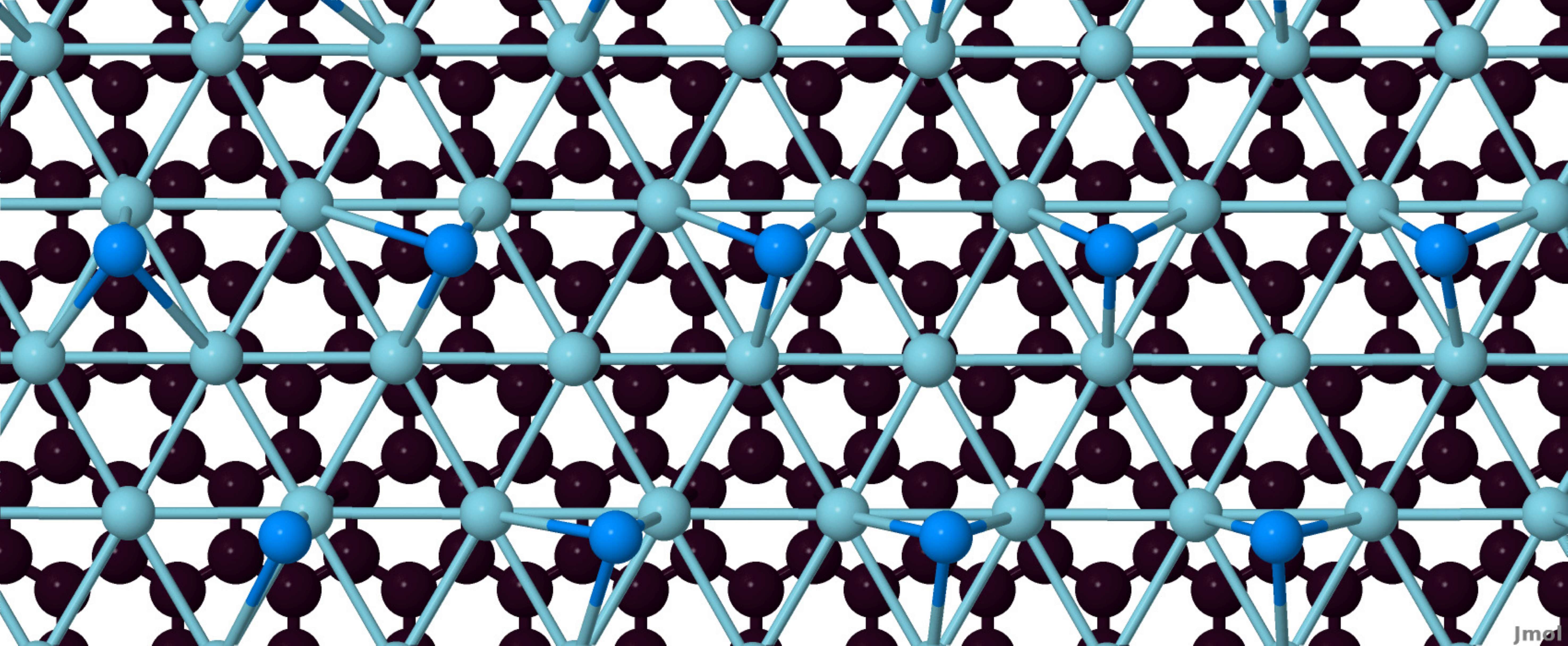}
\caption{\label{Ar-graphite:fig} (Color online)
Successive snapshots of the ``quantized'' sliding state of an Ar
(light-blue/clear) monolayer confined between a static graphite layer
(black), and a nanopatterned top layer (dark blue) advancing rightward at a
speed $v_{\rm ext}= 7.9$~m/s.
The time interval between successive frames is $12.5$~ps.
The top layer lattice spacing $a_{\rm t}\simeq 0.7$~nm, corresponds to
coverage $\Theta=1$ of the antisoliton pattern.
The leftward motion of the Ar layer is evident.
%
}
\end{figure}

To verify this possibility, we simulate this system by adopting the LJ
parameters of the most basic model proposed in Ref.~\onlinecite{Sharma89}.
The main difference with the hiterto studied model is that the bottom
substrate is a honeycomb net, see Fig.~\ref{Ar-graphite:fig}, rather
than the triangular lattice.
For mechanical units we take the graphite in-plane lattice spacing
$a_{\rm b} =a_{\rm graphite} =246.4$~pm,
$m=m_{\rm Ar}=6.63\times 10^{-26}$~kg,
and $\varepsilon_{\rm pp} = \varepsilon_{\rm Ar-Ar} = 10.3$~meV.
The Ar-C interaction energy $\varepsilon_{\rm bp} = \varepsilon_{\rm Ar-C}
= 5.65~{\rm meV}
= 0.549 \, \varepsilon_{\rm Ar-Ar}$ \cite{Sharma89}.
We approximate the Ar lattice constant to $a_{\rm Ar}\simeq 20/13 \, a_{\rm
  graphite}\simeq 379~{\rm pm}$.
For the top substrate we assume a triangular nanopattern with $a_{\rm t} =
20/7\,a_{\rm graphite}\simeq 704$~pm, such to produce a coverage
$\Theta=1$.
For the $\sigma_{\rm tp}$ and $\epsilon_{\rm tp}$ parameters we adopt
tentatively the Ar-C ones \cite{Sharma89}.

Even at the comparably large simulated speed $v_{\rm ext}= 0.05 \, m^{-1/2}
\varepsilon_{\rm pp}^{1/2} = 7.9$~m/s (see Table~\ref{units:tab}), we do
find quantized antisoliton motion, with the lubricant running backward,
precisely at the speed $v_{\rm c.m.}/v_{\rm ext}= w_{\rm quant}=-7/13 \simeq
-0.538$ predicted by Eq.~\eqref{w_quant}.
We verified that the quantized state is retrieved also in the following
conditions: (i) Ar bilayer ($N_{\rm layer}=2$), rather than monolayer, (ii)
the application of $F_{\rm load}=0.004$, representing a 1~MPa load,
and (iii) a looser nanopattern of the top layer, namely
$a_{\rm t} = 40/7\,a_{\rm graphite}\simeq 1408$~pm, i.e.\ $\Theta=2$.
However, we could retrieve no quantized state for other (non-integer)
coverages, at least at the driving speeds we tested.
We conclude therefore that the Ar/graphite system is potentially suitable
for the observation of antisoliton dragging, with a remarkable backward
lubricant motion, provided a nanopatterned top layer of a properly tuned
periodicity can be assembled and brought into contact with the Ar layer.

Analogous incommensurate configurations occur for other noble gases on
metal surfaces such as Ag(111) and Pb(111).
It is quite possible that similar quantized sliding regimes occur in such
systems as well.
However, in some cases the noble gas-metal interaction may be comparably
stronger \cite{Lv11} than with graphite, possibly resulting in a higher
corrugation and practically pinned (anti)solitons.

An experimentally promising geometry which could reveal the quantized
sliding phenomenology could be realized in surface force apparatus (SFA)
experiments \cite{Israelachvili92} where atomically thin lubricant layers
are confined between molecularly smooth mica surfaces.
At a larger (meso) scale, the same mechanism could be realized by some
modification of the setup used in Ref.~\onlinecite{Bohlein12} where a 2D
crystal of colloidal particles is dragged by a flow of solvent over a
periodic corrugation generated by a light interference pattern.
A pattern or solitons or antisolitons, very similar to that illustrated for
an atomic overlayer in Fig.~\ref{modelsketch}, can form and slide around
when the two lattice spacings do not match \cite{Vanossi12PNAS}.
In this case, a second independent periodic interference pattern might be
used to mimic the sliding top layer and drag the soliton pattern along.
At an even larger (macro) scale, friction experiments with a 2D ``granular''
system consisting of photoelastic disks confined in a channel \cite{Krim11}
might be considered with channel walls formed by two corrugated and
vertically oriented Plexiglas sheet, once again reproposing the soliton
mechanisms under shear.

\subsection{The velocity plateau}\label{Conditions:sec}

Quantized sliding, where the ratio $v_{{\rm c.m.}\,x}/v_{\rm ext}$
remains  constant, forming a flat ``plateau'', as a function of
parameters, occurs within certain ranges of physical conditions,
speed, etc.
Of course, plateaus do not extend to arbitrary values of the
physical parameters, but end at certain boundaries marking a sort of
``dynamical phase diagram''.
The point in parameter space where the quantized sliding terminates
identifies a sort of dynamical depinning transition, where the top slider's
grip on solitons is lost.\cite{Vanossi07PRL}.
A variation of system parameters will generally affect the plateau
extension and the precise occurrence of this dynamic depinning.

The most straightforward way to end the quantized sliding state is by
increasing the driving velocity $v_{\rm ext}$.
Indeed, simulations show that, once the plateau exists for a given speed
$v_{\rm ext}$, the quantized state holds for all smaller speeds, at least
at zero or low enough temperature.
In contrast, for increasing $v_{\rm ext}$, beyond a critical speed $v_{\rm
  crit}$ the quantized state is generally lost.
The reason for the existence of such a maximum speed is that the quantized
state is based on the forced advancement of a soliton deformation at speed
$v_{\rm ext}$ along the lubricant crystal.
As soon as $v_{\rm ext}$ is larger than the lubricant speed of sound, the
amplitude of this soliton wave decays rapidly due to inertia, until it
disappears together with the quantized state.
However, whenever the pinning between the soliton pattern and the top
substrate is weak, the depinning may occur earlier, for smaller $v_{\rm
  ext}$.
It is then natural to regard the critical speed $v_{\rm crit}$ as a measure
of the robustness of the quantized state.
We map this robustness under variations of other parameters: temperature
$T$, the load $F_{\rm load}$ per particle in the top layer, the soliton
coverage ratio $\Theta$, and the number of lubricant layers $N_{\rm
  layer}$.

\subsubsection{The quantized state as a function of the driving velocity}
\label{drivingV:sec}

We use sequences of linked MD simulations to investigate the termination of
the quantized-sliding state, as $v_{\rm ext}$ is changed in small steps.
A similar study was carried out for the 1D Frenkel-Kontorova model
\cite{Vanossi04a} and for the 1D and 2D analogous of the present sliding
model \cite{Cesaratto07,Vanossi07Hyst,Manini07PRE}, where a hysteretic
termination of the plateau was identified in underdamped dynamics.
As $v_{\rm ext}$ is increased adiabatically, coming from the low-speed
quantized state, there is a good chance that the ensuing sliding state
remains quantized.
This quantized sliding will therefore generate a plateau of constant
$w=v_{{\rm c.m.}\, x}/v_{\rm ext}$, extending until a critical speed $v_{\rm
  crit}$, where the pinning of solitons to the top slider corrugation
loses its battle against the dissipative forces acting on the lubricant
layer, represented by Eq.~\eqref{eq_Langevin}.
For $v_{\rm ext}\geq v_{\rm crit}$, a non-quantized state ensues,
characterized by an irregular lubricant motion, and a center-mass speed
fluctuating non periodically far from the quantized value $w_{\rm quant}
v_{\rm ext}$.
Upon adiabatic decreasing $v_{\rm ext}$ from this high-speed non-quantized
state, the quantized state is usually recovered at a speed lower than the
depinning $v_{\rm crit}$, a clearly hysteretic unpinning-pinning dynamical
transition.
In the intermediate range, the velocity ratio $w = v_{{\rm c.m.}\, x} /v_{\rm
  ext}$ is therefore a multi-valued function of $v_{\rm ext}$.

\begin{figure}
\centerline{
\includegraphics[width=85mm,angle=0,clip=]{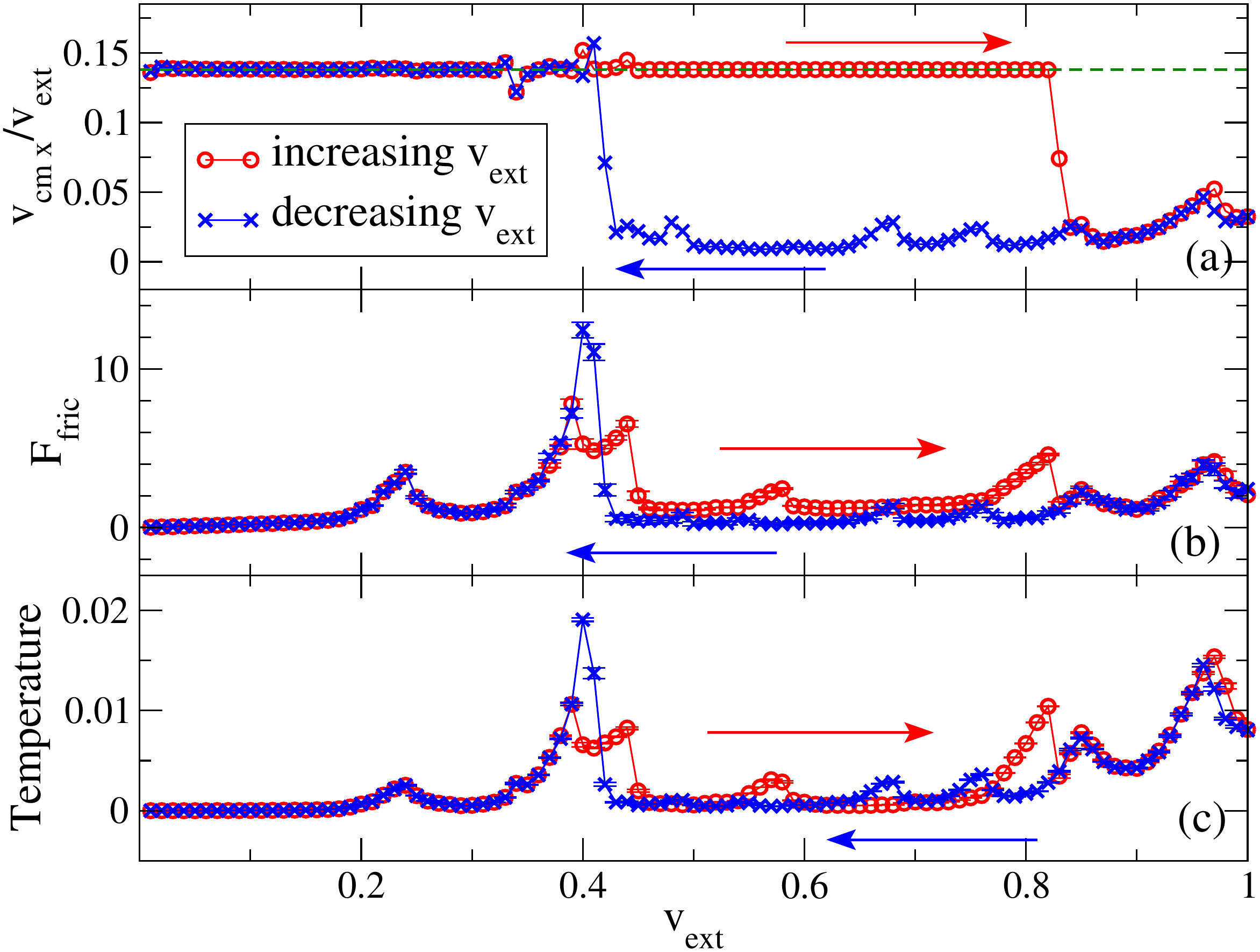}
}
\caption{\label{hysteresis:fig} (Color online)
The plateau of the dynamically pinned state and its tribological
properties, for the model described in Fig.~\ref{modelsketch}.
As a function of the adiabatically increased (circles) or decreased
(squares) top-layer velocity $v_{\rm ext}$, the panels report: (a) the
average velocity ratio $w=v_{{\rm c.m.}\, x}/v_{\rm ext}$ compared to the
plateau value $w_{\rm quant} = 4/29 \simeq 0.1379$, Eq.~\eqref{w_quant},
dashed line; (b) the average friction force experienced by the top layer;
and (c) the average lubricant kinetic energy per particle relative to the
lubricant center of mass.
}
\end{figure}

Figure~\ref{hysteresis:fig}a illustrates this hysteretic depinning for the
fully-commensurate $\Theta = 1$ model of Fig.~\ref{modelsketch}, with
$F_{\rm load} = 0$ and $T = 0$.
The precise value of $v_{\rm crit} $ is obtained by ramping $v_{\rm ext}$
up in small steps; at every step the integration starts from the final
configuration of the preceding step.
For these model parameters, we estimate $v_{\rm crit} = 0.825 \pm 0.005$.
Following the same procedure with downward steps to locate the speed of
recovery of the quantized state, we obtain $v_{{\rm crit}\, \rm down}=
0.415 \pm 0.005$.

The hysteretic loop is due to the ``dynamically metastable'' nature of the
dynamically pinned state.
The finite simulation time $t_{\rm calc}$ and the absence of thermal
fluctuations ($T=0$) can leave the system locked in a dynamically
unfavorable state, which survives until the system jumps into the
appropriate dynamically favored state.

The friction force reported in Fig.~\ref{hysteresis:fig}b, exhibits a
nontrivial structure.
Around $v_{\rm ext}\simeq 0.24$, 0.39, 0.44, 0.58, and 0.82, the friction
force (and consequently the dissipated power) is seen to peak and then drop
to a smaller value.
These friction peaks arise at the resonances of the ``washboard'' frequency
of the advancing lubricant crystal with the bottom lattice with specific
vibrational normal modes of the lubricant lattice.
The resonance are reflected by peaks in the lubricant internal kinetic
energy, see Fig.~\ref{hysteresis:fig}c.
Across these resonant peaks, the value of $w$ remains mostly stable, except
at the last of these transitions, marking the end of the quantized plateau,
with $w$ moving away from the $w_{\rm quant}$ value, again coinciding with
a significant drop in friction.
At resonant peaks rearrangements of the pinned configuration may occur,
with the top layer displacing to grab and drag the soliton pattern to a
different mutual arrangement, always guaranteeing the regular advancement
of the solitons/antisolitons realizing the dynamically pinned state and the
associated quantized velocity.

A similar phenomenon is observed on the way back, decreasing $v_{\rm
  ext}$: The friction force and the lubricant internal kinetic energy
undergo several small jumps corresponding to washboard resonances related
to the top-layer advancement over the non-quantized quasi-static state.
Corresponding to the resonances also $v_{{\rm c.m.}\, x}$ has small bumps,
until eventually the plateau state is recovered, with a sudden jump in the
friction force.
The hysteretic depinning regime observed in the present fully 3D model is
therefore richer than that observed in the purely 1D model
\cite{Manini08Erice} or in the 1+1D model of
Refs.~\onlinecite{Castelli09,Castelli08Lyon}.

If the ideal one-to-one geometrical interlocking between the top
corrugation and the lubricant soliton pattern ($a_{\rm t} = a_{\rm sol}$,
i.e.\ $\Theta= 1$) is of course an especially favorable condition for the
occurrence of dynamical pinning, we do find velocity quantization even for
$\Theta\neq 1$, although not for all investigated values of $\Theta$.
Assuming that the previously unraveled 1D mapping to the Frenkel-Kontorova
model \cite{Vanossi07PRL} is also meaningful in the present richer
interface geometry, the coverage ratio should thus affect the robustness of
the velocity plateau.
Indeed, simulations with simple integer ratios, such as $\Theta=2$ and
$\Theta=1/2$, do show quantized sliding essentially equivalent to the case
with $\Theta=1$.
Other configurations with fractional $\Theta$, where the top-lattice
crystal lines turn to be more pronouncedly out-of-registry with the
lubricant soliton pattern, give rise to a weakening, or even the loss, of
the quantized plateau.

By following the quantized plateau up to its critical speed for several
values of the mismatch ratio $r_{\rm b} = a_{\rm b}/a_{\rm p}$, ranging
from solitonic ($r_{\rm b}>1$) to antisolitonic ($r_{\rm b}<1$), we find a
rather erratic dependence of $v_{\rm crit}$ on $r_{\rm b}$.
In this case, the different degree of efficiency of the grip on solitons
and thus of robustness of the quantized dynamics, may be partially related
to random initial conditions, hardly a controllable element.

\subsubsection{Effects of temperature}\label{temperature:sec}

\begin{figure}
\centerline{
\includegraphics[width=85mm,angle=0,clip=]{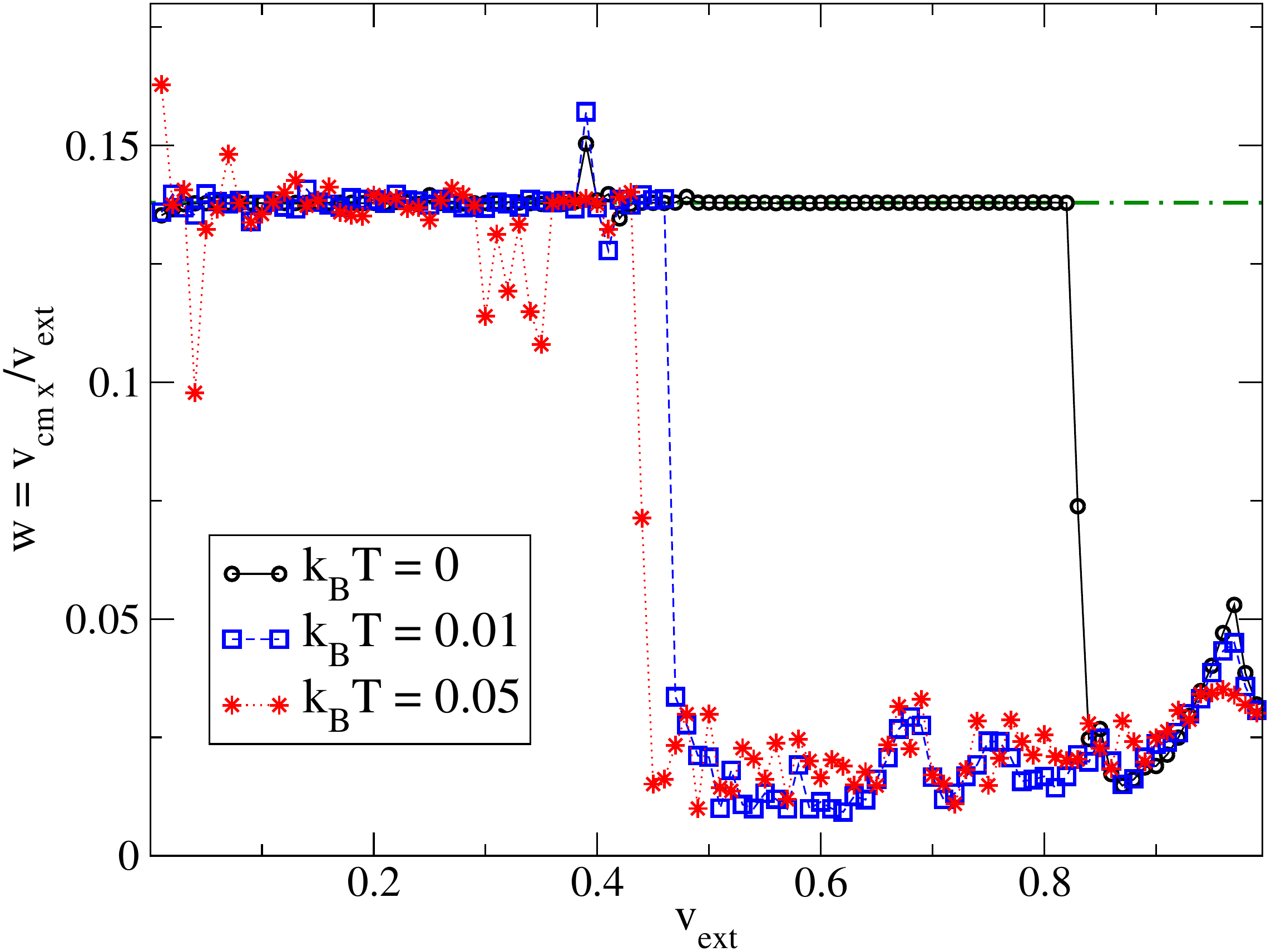}
}
\caption{\label{temperature:fig} (Color online)
The effect of temperature on the plateau of the dynamically pinned state
studied for increasing $v_{\rm ext}$.
The circles are the same $T=0$ data as in Fig.~\ref{hysteresis:fig}a.
Each square or star is obtained as by averaging $v_{{\rm c.m.}\,x}$ over
the last 70\% of an at least 100 time units long Langevin simulation at
finite temperature, started from the final state of the previous
configuration.
The dot-dashed line marks the plateau value $w_{\rm quant} = 4/29$ for the
considered geometry.
}
\end{figure}

To investigate the robustness of the quantized state against thermal
fluctuations, we run finite-temperature simulations in the same conditions
as the zero-temperature runs discussed until now.
The results are summarized in Fig.~\ref{temperature:fig}.
For low temperature $k_{\rm B}T=0.0001$ and $0.001$ (not shown), even
though the trajectories of individual particles are affected by thermal
fluctuations, $w$ exhibits no significant deviation from $T=0$.

For larger $k_{\rm B}T=0.01$ and $0.05$ we observe deviations and
fluctuations around the quantized plateau speed, see
Fig.~\ref{temperature:fig}.
Similar deviations were found in the 1+1D model \cite{Castelli09,
  Castelli08Lyon}.
Notice that these deviations in $w$ reflect very wide instantaneous
fluctuations, often far exceeding the average lubricant velocity.
The averaging over a finite simulation duration $t_{\rm calc}$ integrates
out these large fluctuations, suggesting that, over an appropriately
reduced range of $v_{\rm ext}$, the system fluctuates around the
quantized sliding state, which still dictates the average lubricant
advancement speed.
If longer simulations were carried out, further averaging would decrease
the fluctuation amplitude, thus indicating that the thermal regime is
indeed randomly fluctuating around the dynamically pinned state.
In temperature, the dynamical depinning tends to occur at a generally
smaller driving speed.

For even larger $k_{\rm B}T=0.1$, the tendency to in-plane thermal
expansion of the lubricant layer, frustrated by the in-plane PBC, resolves
in the expulsion of a small fraction of atoms from the lubricant layer,
which thus gets rid of the soliton-originating mismatch to the bottom
layer.
As a result, the quantized sliding state is completely absent at such high
temperature.

\subsubsection{Effects of applied load}\label{Fload:sec}

\begin{figure}
\centerline{
\includegraphics[width=85mm,angle=0,clip=]{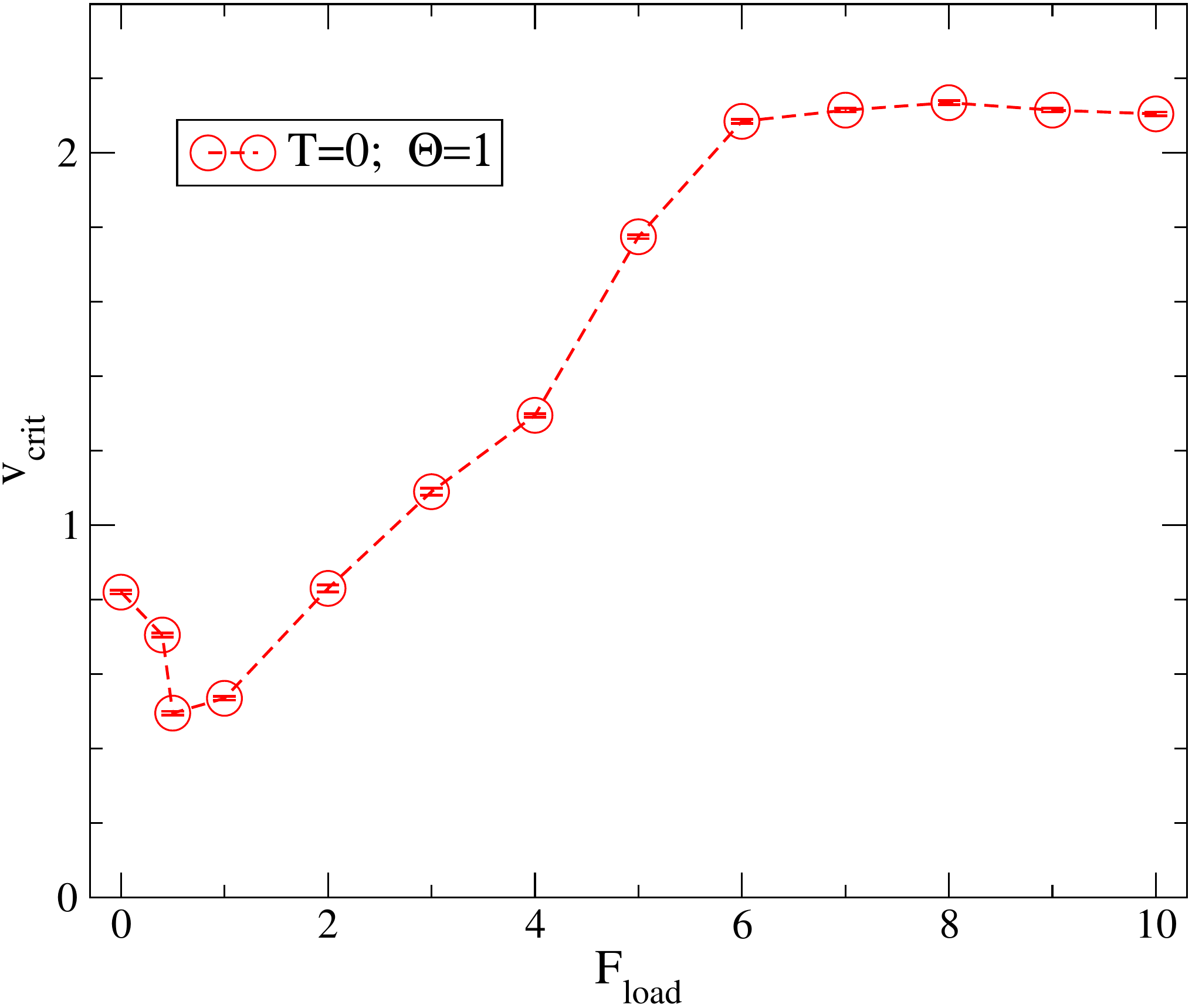}
}
\caption{\label{load:fig} (Color online)
The depinning speed $v_{\rm crit}$ as a function of the applied load per
particle $F_{\rm load}$ for the same model as in Fig.~\ref{modelsketch}.
Beyond the small-load region, the quantized-sliding state exhibits an
overall benefit of increased load.
}
\end{figure}

We also investigate the effect of changing the load applied between the
sliders, squeezing the lubricant layer among them.
The lubricant in turn is not perfectly flat, because in-register regions
are composed by hollow-site atoms, which move closer to the bottom slider,
while the soliton regions consist of atoms occupying bridge or top sites,
which are therefore pushed upward.
In matched ($\Theta=1$) configurations and in the ensuing quantized sliding
state, the top slider atoms tend to catch over the in-register regions
which are the most vertically depressed lubricant areas, rather than over
the solitons, where the lubricant is sticking out locally.
As a result, the applied load squeezes down onto the in-register regions,
and affects the solitonic regions more marginally. 
Thus, the increased load should make it more difficult for the soliton
pattern to unpin itself from the top-layer corrugation.

To investigate the load dependence of the quantized plateaus we consider
several $F_{\rm load}$ values, and for each of them we cycle $v_{\rm ext}$
up in small steps, as described in Sect.~\ref{drivingV:sec}, to determine
$v_{\rm crit}$.
We collect the resulting values of $v_{\rm crit}$ for varied load in
Fig.~\ref{load:fig}, which shows that, by increasing $F_{\rm load}$,
$v_{\rm crit}$ generally rises, thus indicating that, as expected, the
quantized state is extended under a larger load $F_{\rm load}$.

\subsubsection{Multiple lubricant layers}

\begin{figure}
\centerline{(a)\hfill
\includegraphics[width=80mm,angle=0,clip=]{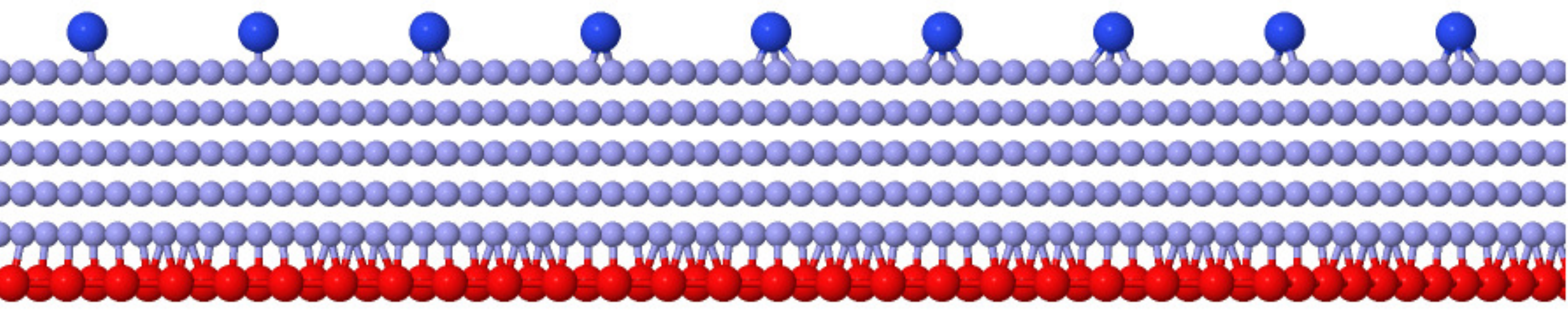}
}
\centerline{(b)\hfill
\includegraphics[width=80mm,angle=0,clip=]{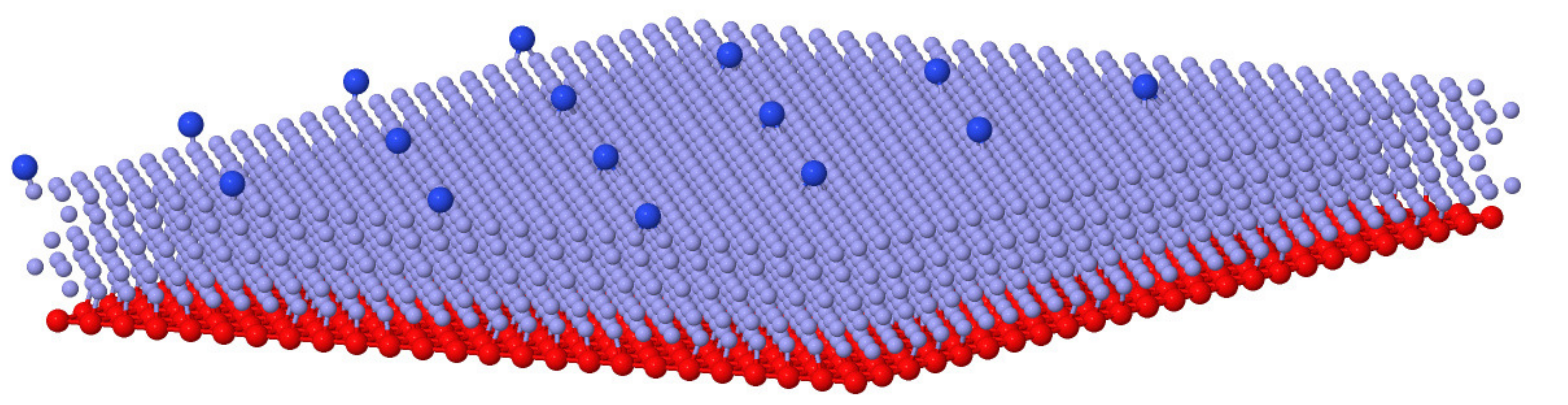}
}
\caption{\label{5layers:fig} (Color online)
(a) Side and (b) perspective view of a $N_{\rm layer}=5$ lubricant layers
model, with the same lattice mismatch and other parameters as in
Fig.~\ref{modelsketch}.
}
\end{figure}

\begin{figure}
\centerline{
\includegraphics[width=85mm,angle=0,clip=]{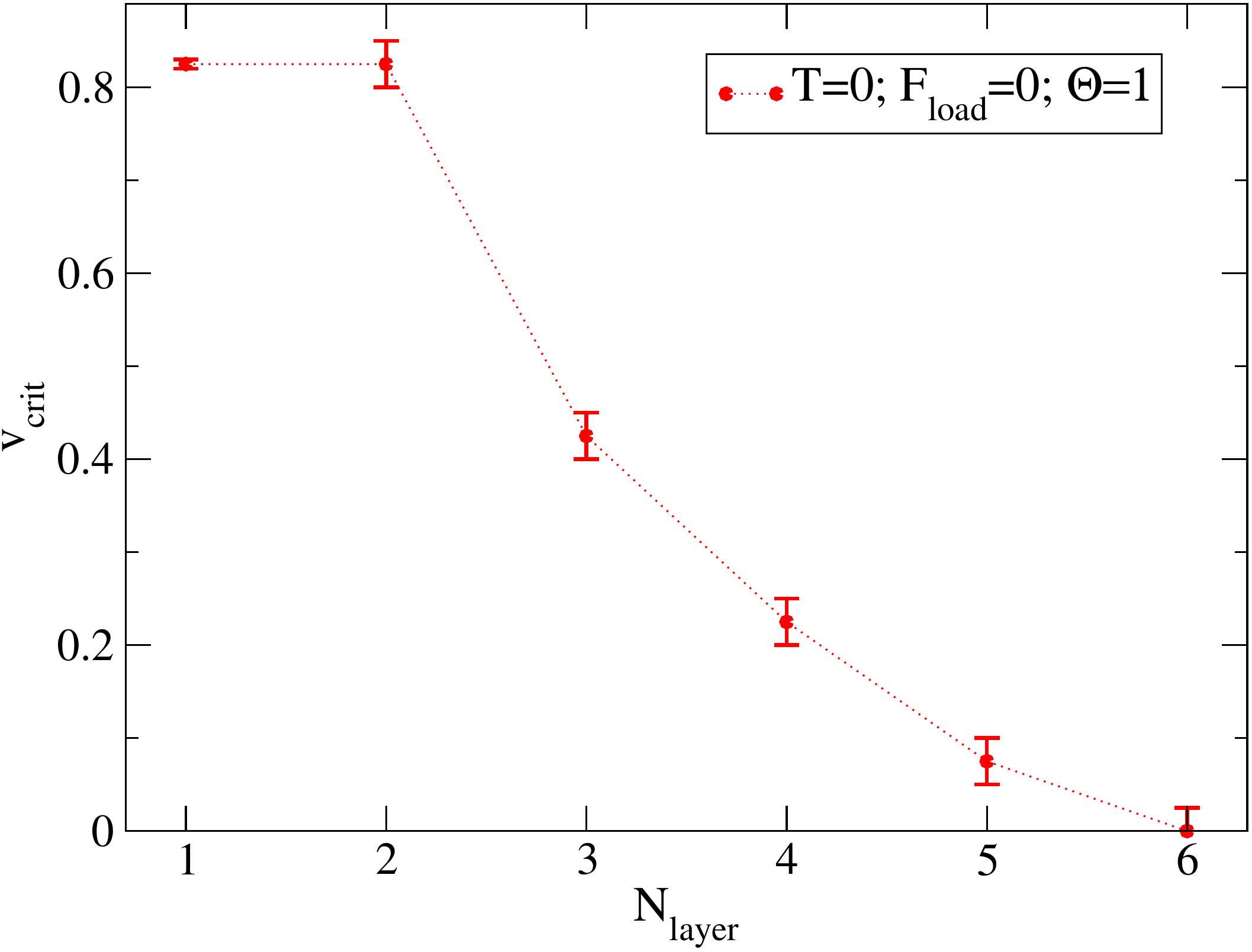}
}
\caption{\label{multilayers:fig} (Color online)
The depinning speed $v_{\rm crit}$ as a function of $N_{\rm layer}$ for a
multilayer configuration of the type illustrated by Fig.~\ref{5layers:fig},
with the same simulation parameters as in Fig.~\ref{modelsketch}.
The quantized sliding state weakens for increasing number of lubricant
layers $N_{\rm layer}\geq 2$.
}
\end{figure}

In boundary lubrication, the lubricant as a rule solidifies into a
multiplicity of layers, whose thickness is gradually reduced by squeeze-out
under pressure, until a single layer is just an extreme possibility.
It is therefore important to verify whether the plateau dynamics is an
exclusive prerogative of the single lubricant layer studied so far, or
whether it will occur even for multilayer solid lubricant films -- although
of course with generally smaller and less robust plateaus.
Figure~\ref{5layers:fig} displays the typical arrangement of lubricant
particles relative to the substrates in a lubricant multi-layer
configuration.
Soliton deformation affect mostly the lubricant layer in direct contact
with the bottom substrate.
The atoms of the uppermost lubricant layer are spaced almost regularly, but
the residual vertical displacements can be sufficient for the soliton
pattern to ingrain in the top substrate.

For multiple lubricant layers (up to $N_{\rm layer} = 5$), we recover
quantized velocity plateaus, for the case examined of full matching
$\Theta=1$, with $F_{\rm load}=0$ and $T=0$.
We evaluate the robustness of the quantized sliding state by determining
the critical speed $v_{\rm crit}$ where the quantized plateau ends in this
multilayer lubricant case.
Figure\ \ref{multilayers:fig} shows that the broadest plateau is
achieved for $N_{\rm layer} = 1$.
Its width is still as large at $N_{\rm layer} =2$; further lubricant
thickening reduces $v_{\rm crit}$ progressively.
This decrease is not surprising, as the power-law weakening of
soliton-induced corrugation across the film makes the grip on solitons by
the top slider harder and harder for thicker and thicker layers.
For $N_{\rm layer}>5 $, we could detect no quantized-sliding dynamics, even
at very small $v_{\rm ext}$.

\section{Discussion and conclusion}

We present a simulation study of the relative sliding of rigid
incommensurate crystal surfaces separated by a 3D solid and fully mobile
lubricant film, whose interatomic interactions were assumed to be of LJ
type.
The ``quantization'' of the lubricant's sliding speed previously uncovered in
much more idealized, lower dimensional models is fully confirmed in this
more realistic case.
The quantized relative speed plateau as a function of overall sliding speed
is detected very clearly and demonstrated to extend over broad parameters
ranges including applied load, number of lubricant layers and
commensuration ratio between the top layer and the soliton lattice.

Focusing mainly on unrotated lattices and a single lubricant layer, we find
perfect plateaus at the same geometrically determined velocity ratio
$w_{\rm quant}$ as observed in the 1D and 2D models, both in case of
solitons (forward lubricant sliding) and of antisolitons (backward soliton
sliding).
We find that the soliton pinning to the top slider leading to plateau
quantization is abandoned by increasing the sliding velocity $v_{\rm ext}$
above a critical value $v_{\rm crit}$.
It is eventually retrieved when $v_{\rm ext}$ is reduced back down to
$v_{{\rm crit}\, \rm down}< v_{\rm crit}$, thus with a hysteresis.
The quantized sliding state is strengthened by an applied load.
Although the optimal rate of commensuration for quantization to occur is
perfect 1:1 matching ($\Theta =1$) between soliton lattice and top slider
lattice of kinks to the upper slider lattice, weaker but definite quantized
regimes exist even for $\Theta \neq 1$ .

In the attempt to address slightly more realistic conditions, we also model
a multilayer as opposed to monolayer LJ solid lubricant; and a monolayer
and a bilayer of solid Ar acting as a lubricant between a flat graphite
surface and a nanopatterned slider.
Quantized sliding is recovered in both cases, although in a rather fragile
form for $N_{\rm layer} > 3$.
We see no reason for the same Moir\'e-pattern dragging mechanism to be
restricted to LJ systems: It is likely to show up in many sliding-friction
experiments, as long as a crystalline lubricant thin film (e.g.\ a graphene
layer) is sandwiched in between two different lattice-mismatched
crystalline sliders.

The present preliminary investigation of thermal effects confirms the
robustness of the quantized state.
Like for the 1+1D model of Refs.~\onlinecite{Castelli09,Castelli08Lyon}, we
find that (i) the quantized plateau becomes noisy, with the relative
lubricant velocity $w$ fluctuating around $w_{\rm quant}$, (ii) the
dynamical depinning, rather than a sharp hysteretic transition, behaves as
a continuous crossover, and (iii) this crossover occurs at a generally
smaller speed $v_{\rm ext}$.
A further systematic investigation of thermal effects and of the mutual
rotation of the three crystalline layers promises nontrivial developments.


\acknowledgments

We acknowledge useful discussion with P. Ballone and I.E. Castelli.
This work was partly supported by ERC Advanced Research Grant N. 320796
MODPHYSFRICT, by MIUR, through PRIN-2010LLKJBX-001, by SNSF, through
SINERGIA Project CRSII2 136287/1, by COST Action MP1303,
and by the EU-Japan Project LEMSUPER.



\end{document}